\documentclass[10pt,aps,prc,floatfix,twocolumn,nofootinbib, superscriptaddress,preprintnumbers]{revtex4-1}

\usepackage{amsfonts,amsmath,amssymb,bm, xspace}
\usepackage{color,graphicx}
\usepackage{bbm}
\usepackage{dcolumn}                 
\graphicspath{{./figures/}}
\usepackage{graphicx}
\usepackage{float}
\usepackage[dvipsnames]{xcolor}
\usepackage{bbold}
\usepackage{enumitem}
\usepackage{subcaption} 
\usepackage{caption} 
\usepackage[colorlinks=true,
linkcolor=blue,anchorcolor=blue,citecolor=blue,
hypertexnames=false]{hyperref}                

\usepackage{array,physics}
\usepackage[normalem]{ulem}
\usepackage{tikz}
\newcolumntype{d}[1]{D{.}{.}{#1}}

\newcommand{\sat}{\textrm{sat}}
\newcommand{\sym}{\textrm{sym}}
\newcommand{\nsat}{n_\sat}
\newcommand{\NM}{\textrm{NM}}

\renewcommand{\max}{\textrm{max}}
\newcommand{\obs}{\textrm{obs}}
\newcommand{\tov}{\textrm{TOV}}
\newcommand{\msun}{M_\odot}

\hypersetup{%
    pdfsubject=Paper,
    pdfkeywords={nuclear physics} 
    unicode=true,
    breaklinks=true,
     colorlinks   = true,
     linkcolor = blue,
     citecolor = blue,
     menucolor = blue,
     urlcolor = blue
}

\begin{document}

\title{Relativistic mean-field predictions for dense matter equation of state and application to neutron stars}

\author{Luca Passarella}
\email{luca\_passarella@polito.it}
\affiliation{Department of Applied Science and Technology, Politecnico di Torino, I-10129, Torino, Italy}
\affiliation{INFN, Sezione di Torino, I-10126, Torino, Italy}
\author{J\'er\^ome Margueron}
\affiliation{International Research Laboratory on Nuclear Physics and Astrophysics, Michigan State University and CNRS, East Lansing, MI 48824, USA} 

\author{Giuseppe Pagliara}

\affiliation{Department of Physics and Earth Science, University of Ferrara, via Saragat 1, I-44122 Ferrara, Italy}
\affiliation{INFN, Sezione di Ferrara, via Saragat 1, I-44122 Ferrara, Italy}
\date{\today}

\begin{abstract}
Relativistic mean-field models (RMF) based on the exchange of $\sigma$, $\omega$, and $\rho$ mesons including non-linear nucleon-$\sigma$ couplings and density-dependent $\rho$ coupling, are considered. A large set of models is generated using the Markov chain Monte Carlo approach and Bayesian statistics to reproduce nuclear physics knowledge encoded in terms of the nuclear empirical parameters and $\chi$EFT predictions for low-density neutron matter. These models are filtered, in a second step, using astrophysical constraints: the tidal deformability obtained from GW170817 parameter estimation and the observational masses deduced from radio-astronomy. We then obtain a set of selected RMF models that are compatible with present nuclear and astrophysical constraints and that can be employed to make predictions and to quantity their uncertainties. Predictions for masses and radii are compared to NICER masses-radii analyses for PSR J0030+0451 and PSR J0740+6620. We find that RMF models can be made soft enough to predict low values for neutron star radii compatible with GW170817 and, at larger densities, stiff enough to be compatible with NICER analyses for massive neutron stars. Our models can also reach large values for the maximum mass, up to 2.6$M_\odot$. 
In addition, for the core composition, we obtain a large distribution of the proton fraction for canonical mass neutron stars, some of them allowing the direct URCA fast cooling process. For massive neutron stars, however, most of our models suggest a large proton fraction in the core allowing direct URCA fast cooling process. 

\end{abstract}

\maketitle

\section{Introduction}

The determination of the equation of state (EOS), including uncertainty quantification, for cold and dense nuclear matter represents one of the most intriguing and complex questions in nuclear physics. In particular, while the properties of nuclear matter are rather well understood at densities close to nuclear saturation density $n_{\sat}\approx 0.155$~fm$^{-3}$ (corresponding to the energy-density $\rho_\sat\approx 2.7\times 10^{14}$~g~cm$^{-3}$) and at small isospin asymmetries ($Z/2\lesssim N \lesssim 2Z$, where $N$ and $Z$ are the numbers of neutrons and protons in a unit volume) owing to the large amount of data on finite nuclei issued from nuclear physics laboratories, extrapolations to supra-saturation densities and large isospin asymmetries are quite uncertain due to many-body intrinsic difficulties. The deep knowledge of the EOS would, however, allow us to understand the properties of some of the most fascinating astrophysical stellar objects, namely Neutron Stars (NSs). 

In the literature, two standard approaches have been used: ab-initio calculations, see e.g. Refs.~\cite{Baldo:1997ag,Bogner:2009bt,Carlson:2014vla}, and phenomenological models, see e.g. Refs.~\cite{Mueller:1996pm,Margueron:2009jf}, both having their advantages and drawbacks. In the first case, the properties of nuclear matter can be directly related to the properties of the nucleon-nucleon interactions (also including three-body forces). Among the several ab initio calculations, a very powerful approach is the one of chiral effective field theory, $\chi$EFT. The main advantage of this approach is its systematic expansion in powers of nucleon momenta and its connection to field theory. As a consequence, the renormalization of the approach is performed at each order in the expansion. Moreover, the errors deriving from the truncation of the expansion and the statistical uncertainties from the fitting data are quantified. 
Apart from its large computational effort, calculations within $\chi$EFT are reliable up to densities of, at most, twice $n_{\sat}$ thus not large enough to model the central regions of NSs. Phenomenological models instead make use of the Nuclear Empirical Parameters (NEP) to fix the couplings between nucleons and exchange mesons and the EOS is usually computed in the mean field approximation. The major advantage of this approach is the low computational effort that allows the exploration of wide ranges of density, isospin asymmetries, and also temperatures. Major drawback: there is no guideline for introducing interaction terms in the underlying Lagrangian, terms (and corresponding couplings) which are therefore purely phenomenological and whose contribution, especially at large densities, is not fully under control.

Once the EOS is obtained within these theoretical frameworks, the structure of NSs (as for instance the mass-radius relations) can be computed and compared with observational results. This procedure however is solely based on the physics close to $n_{\sat}$ and does not exploit the information on the EOS that can be extracted from the observed properties of NSs. In the seminal paper~\cite{Steiner:2010fz}, for the first time, nuclear physics input has been integrated with astrophysical inputs through a Bayesian analysis of the astrophysical data, and the EOS was reconstructed by use of a polytropic parametrization at high-density. Since then, several papers have used a similar approach, see for instance Refs.~\cite{Margueron:2017lup,Margueron:2017eqc,Drischler:2020hwi,Traversi:2020aaa,Alvarez-Castillo:2020,Malik:2022zol,Zhu:2023,Beznogov:2023} for studies on nucleonic matter. Alternatively, also approaches based on neural networks have been recently applied to address the same problem, see e.g. Ref.~\cite{Guo:2023mhf}.

Here, we resort to Bayesian analyses and make use of a large set of relativistic mean field models to infer the EOS of dense matter by exploiting the experimental knowledge of the NEP, the theoretical constraints on neutron matter arising from ab initio predictions based on $\chi$EFT, the astrophysical constraints on the maximum mass of NSs and the tidal deformability associated with the gravitational waves signals GW170817. Finally, our EOS inference is compared to the recent constraints on the masses and radii of PSR J0030+0451 and PSR J0740+6620 as obtained by the analyses of NICER observatory data. Here, we limit our study to the case of matter containing nucleons and leptons and disregard the possibility of the formation of hyperons and delta resonances, see Ref.~\cite{Drago:2014oja,Malik:2022jqc,Sun:2023} for detailed studies on their formation in dense matter and \cite{Suprovo:2022,Huang:2024rvj,Parmar:2025csx} for recent Bayesian analyses on hyperonic matter. We show in particular that it is possible to find parametrization of the EOS that satisfies the nuclear physics and astrophysics data under consideration. Moreover, we obtain predictions for the range of the maximum mass of NSs, the central baryon densities of most massive objects, the proton fraction in the core of NSs, and the size of the region that undergoes fast cooling for two representative masses, $1.4M_\odot$ and $2.0M_\odot$.

\section{Relativistic mean field models}

A version of the relativistic mean field model devised in Refs.~\cite{PhysRevLett.67.2414, Glendenning2000} is employed in a Markov chain Monte-Carlo approach combined with the Bayesian statistics. In this model, the interaction between nucleons is mediated by the exchange of  $\sigma$, $\omega$, $\rho$ mesons, and non-linear self-couplings of the $\sigma$ field are also adjusted to reproduce low values of the nuclear matter incompressibility. In addition, the coupling between nucleons and the $\rho$ meson, $g_\rho$, is being changed into a density-dependent coupling $g_{\rho}(n_b)$, as suggested for instance in Ref.~\cite{Typel:1999yq}. The corresponding Lagrangian reads:
\begin{equation}\begin{split}
\label{Glen eq 4.219}
\mathcal{L}= \sum_{N} \bar{\psi}_{N} ( i \gamma _{\mu}\partial ^{\mu} - m_{N} + g_\sigma\sigma
-g_\omega \gamma _{\mu} \omega^{\mu} \\
-\frac{1}{2}g_{\rho}(n_b)\, \gamma _{\mu} \boldsymbol{\tau} \cdot \boldsymbol{\rho} ^ {\mu})\psi_{N}
+\frac{1}{2}( \partial_{\mu} \sigma \partial^{\mu} - m^2 _{\sigma} \sigma ^2) \\
-\frac{1}{4} \omega _{\mu \nu} \omega ^{\mu \nu} + \frac{1}{2} m^2 _{\omega} \omega _{\mu} \omega ^{\mu} 
-\frac{1}{4} \boldsymbol{\rho} _{\mu \nu} \cdot \boldsymbol{\rho} ^{\mu \nu} \\    -\frac{1}{2} m_{\rho}^2 \boldsymbol{\rho} _{\mu} \cdot \boldsymbol{\rho} ^{\mu}
-\frac{1}{3}b m_N (g_{\sigma} \sigma)^3 -\frac{1}{4}c(g_{\sigma} \sigma )^4
\end{split} 
\end{equation}
where $\psi_{N}$ is the nucleon $N$ spinor ($N$ can be a neutron or a proton). 

Anticipating the discussion of the results, it is interesting to compare the Lagrangian we consider~\eqref{Glen eq 4.219} with the one considered in a recent analysis in Ref.~\cite{CHuang:2024}. The main differences are the following ones: the meson couplings between $\rho$ and $\omega$ as well as the non-linear $\omega$ coupling are considered in Ref.~\cite{CHuang:2024} while they are absent in our Lagrangian~\eqref{Glen eq 4.219}. We, however, consider a density-dependent coupling for $\rho$ meson with nucleons, see Eq.~\eqref{eq:grho} in the following. These differences in our approaches impact the nuclear equation of state in asymmetric matter since in the two Lagrangians, coupling constants associated with $\rho$ meson are introduced, but in a different way. 
There are, however, similarities in the prediction of the two approaches, e.g., they both predict large mass for stable NSs, but there are also differences, e.g., the compactness of NSs can be larger with the Lagrangian we consider~\eqref{Glen eq 4.219} compared to the predictions from in Ref.~\cite{CHuang:2024}. These differences are discussed in Sec.~\ref{sec:astro:nicer}.

Let us introduce the model parameters and how they are constrained. There are five couplings constants, namely $g_{\sigma}$, $g_{\omega}$, $g_{\rho}$, $b$, $c$, and one more parameter $a_{\rho}$ describing the density-dependent coupling of $g_{\rho}(n_b)$, where $n_b=\sum_N n_N=n_n+n_p$, as:
\begin{equation}
g_{\rho}(n_b) = g_\rho(\nsat) \exp\Big[- a_\rho \Big(\frac{n_b}{n_{\sat}}- 1\Big)  \Big]\,,
\label{eq:grho}
\end{equation}
as in Ref.~\cite{PhysRevC.104.015802}. The value $g_\rho(\nsat)$ is a constant that is adjusted as explained hereafter.

The density dependence of $g_{\rho}$ introduces a rearrangement contribution $\Sigma_r$ to the nucleon self-energies:
\begin{equation}
\Sigma_r = -\frac{1}{2}a_{\rho}g_{\rho}(n_b)\frac{n_p-n_n}{n_{\sat}}\rho_{03}
\end{equation}
where $\rho_{03}$ is the mean field isospin 3-component of the $\rho$ field (see discussion in the following). The density-dependence of the $\rho$ meson coupling is mostly correlated with the slope of the symmetry energy $L_\sym$.

\begin{table*}[t]
\centering
\tabcolsep=0.5cm
\def\arraystretch{1.4}
\begin{tabular}{cccccccccc}
\hline
\textbf{NEPs} & $E_{\sat}$ & $n_{\sat}$ & $K_{\sat}$ & $E_{\sym,2}$ & $L_{\sym}$ & $m^*_{D,\sat}$ \\
\hline
unif.dist. & $[-16.4,-15.2]$ & $[0.145,0.165]$ & $[180,280]$ & $[28,36]$ & $[30,90]$ & $[0.45,1.05]$ \\
+$\chi$EFT band & $-15.78^{+0.40}_{-0.42}$ & $0.16^{+0.01}_{-0.01}$ & $228.44^{+34.25}_{-33.27}$ & $32.46^{+1.76}_{-1.63}$ & $60.10^{+15.23}_{-13.21}$ & $0.72^{+0.12}_{-0.15}$ \\
+GW170817 & $-15.77^{+0.36}_{-0.38}$ & $0.16^{+0.01}_{-0.01}$ & $231.19^{+30.10}_{-31.91}$ & $31.84^{+1.06}_{-1.09}$ & $51.97^{+7.08}_{-8.19}$ & $0.75^{+0.08}_{-0.08}$ \\
+$M_\tov$ PDF & $-15.77^{+0.36}_{-0.38}$ & $0.16^{+0.01}_{-0.01}$ & $231.49^{+29.99}_{-32.23}$ & $31.83^{+1.07}_{-1.00}$ & $53.44^{+6.06}_{-6.69}$ & $0.70^{+0.05}_{-0.05}$ \\
\hline
\end{tabular}%
\caption{Range of Nuclear Empirical Parameters (NEPs) considered in the present study. In the second, third and fourth row, the centroidal values, with a standard deviation of $\pm$ 1 sigma, are reported for the NEPs displayed in the corner plots of Fig. \ref{fig:corner_plot_tidal} and Fig. \ref{fig:comparison_plot_2_3}.
Each raw show the impact of additional constraints which are described in the text.
}
\label{tab:nep_corrected_version}
\end{table*}

By minimizing the action deduced from Eq.~\eqref{Glen eq 4.219} it is possible to obtain the Euler-Lagrange equation:
\begin{equation}
\begin{split}
\label{Glen eq 4.220}
[ (\gamma _{\mu}(k^{\mu} - g_{\omega}\omega^{\mu} -\frac{1}{2}g_\rho(n_b) \gamma _{\mu} \boldsymbol{\tau} \cdot \boldsymbol{\rho} ^ {\mu}) - m_{D}^*] \psi_{N} (k)=0\, ,
\end{split} \end{equation}
where the Dirac effective mass is defined as,
\begin{equation}
m_{D}^*(\sigma) = m_N -g_\sigma \sigma  \, .
\end{equation}
The eigenvalues for nucleonic states are functions of the momentum $k$ as,
\begin{subequations}
\begin{align}
e_{N}(k)&= g_\omega\omega_0 + g_\rho\rho_{03}\boldsymbol{I}_{3N} + \Sigma_r+ \sqrt{k^2 + (m_{D}^*)^2}\, ,
\end{align}
\end{subequations}

where $\boldsymbol{I}_{3N}$ is the isospin 3-component for nucleons. In uniform matter at static equilibrium, space and time derivatives can be dropped, and the meson field equations are
\begin{subequations}
\begin{align}
\omega_0 &= \sum_N\frac{g_\omega}{m^2_{\omega}} n_N, \\
\rho_{03} &= \sum_N\frac{g_\rho}{m^2_{\rho}}\boldsymbol{ I}_{3N} n_N, \\
g_{\sigma}\sigma &= \frac{g_{\sigma}^2}{m^2_{\sigma}}\Big[-b m_N (g_{\sigma}\sigma)^2 - c (g_{\sigma}\sigma)^3 + \sum_Nn_{sN} \Big], \label{eq:sigma}
\end{align}
\end{subequations}
with the scalar density for nucleon $N$ defined as 
\begin{equation}
n_{sN} = \frac{(g=2)}{2\pi^2}  \int_{0}^{k_{F_N}} k^2 dk \frac{m_{D}^*}{\sqrt{k^2 + (m_{D}^*)^2}}  \, .
\label{eq:scalar_density_hyperons}
\end{equation}
This integral can be obtained analytically and we have
\begin{equation}
n_{sN} = \frac{1}{2\pi^2} m_{D}^* \Biggl[k_{F_N} E_{F_N}+m_{D}^{*2} \ln \Biggl(\frac{m_{D}^*}{E_{F_N}+k_{F_N}} \Biggr) \Biggr] \, ,
\label{eq:scalar_density_2}
\end{equation}
with $E_F=\sqrt{k_F^2+m_{D}^{*2}}$.

The energy density is
\begin{equation}\begin{split}
\epsilon = \frac{1}{3} b m_N (g_{\sigma} \sigma)^3 + \frac{1}{4}c (g_{\sigma}\sigma)^4 +\frac{1}{2}m_{\sigma}^2 \sigma ^2       +      \frac{1}{2}m_{\omega}^2 \omega _{0}^2 \\
+\frac{1}{2}m_{\rho}^2 \rho_{03} ^2 + \frac{g}{2\pi^2} \sum_N \Biggl[   \int_{0}^{k_{F_N}} k^2 dk \sqrt{k^2+(m_{D}^*)^2} \Biggr]\label{eq:en dens}
\end{split}
\end{equation}
while the pressure
\begin{equation}
\begin{split}
p = - \frac{1}{3} b m_N (g_{\sigma} \sigma)^3 - \frac{1}{4}c (g_{\sigma}\sigma)^4 -\frac{1}{2}m_{\sigma}^2 \sigma ^2 + \frac{1}{2}m_{\omega}^2 \omega _{0}^2 \\
+\frac{1}{2}m_{\rho}^2 \rho_{03} ^2+n_b\Sigma_r+\frac{g}{6\pi^2} \sum_N \Biggl[ \int_{0}^{k_{F_N}} \!\!\!\frac{k^4 dk }{{\sqrt{k^2+ (m_{D}^*)^2} }} \Biggr] \, .
\end{split}
\end{equation}

Let us now introduce the NEP, we are going to use to constrain those six parameters: $E_{\sat}$ is the binding energy per nucleon, $n_{\sat}$ is the saturation density, $K_{\sat}$ is the compression modulus, $E_{\sym,2}$ is the quadratic contribution to the symmetry energy, $L_{\sym}$ is the slope of the symmetry energy, and $m^*_{D,\sat}$ is the nucleon Dirac mass (in units of $m_N$). All those quantities are defined at saturation and for symmetric matter. These NEP are known with given uncertainties, which can be large in some cases. The values for the NEP that we consider in our parameter exploration are given in Tab.~\ref{tab:nep_corrected_version} and their relations with the six Lagrangian parameters are given in the appendix. 

Some sets of parameters create models that are found, however, to be in contradiction with NS observations. For instance, some models may predict NS configurations with a maximum mass below the observational limit of $\approx 2M_\odot$. These models have to be removed from our selection. For this reason, we employ in the following astrophysical data to further constrain the values of the Lagrangian parameters. 
In addition, the selected models are used to make predictions on several observables that are not employed for the selection (by the lack of astrophysical data to compare with). Employing selected models for such predictions informs us about the efficiency of the astrophysical data that we have considered to constrain other quantities, e.g., core composition or crust thickness, for which there are no data. The dispersion between the selected models for these quantities provides an uncertainty quantification.  

The EOSs we compute correspond to two cases:  i) neutron matter for the comparison with the theoretical predictions based on $\chi$EFT results and ii) nucleonic matter in beta-equilibrium (with the contribution of leptons and with the additional condition of charge neutrality) for the comparison with the astrophysical data. In the last case,
the EOS for the core has to be completed with EOS for the crust of NSs.
As customary, we use the Baym-Pethick-Sutherland (BPS) EOS for densities smaller than $\sim 0.08$ fm$^{-3}$~\cite{Baym:1971pw}. 
The EOSs with crust and core components are the input ingredient for solving the Tolman–Oppenheimer–Volkoff equations for the structure of non-rotating NSs. In particular, besides the mass-radius relations, we will compute also the tidal deformabilites for comparing with the constraints obtained on this quantity by the observation of GW170817 \cite{Abbott:2017}.

\section{Neutron stars}

We briefly describe how neutron star properties are determined from general relativity equations. 

\subsection{TOV and pulsation equations}

The masses, radii and pressure profiles of non-rotating NSs are obtained by solving the Tolman–Oppenheimer–Volkoff (TOV) equations:
\begin{equation}
\begin{cases}
\displaystyle \frac{dm}{dr} = 4\pi r^2 \epsilon(r) \\[15pt] 
\displaystyle \frac{dp}{dr} = \frac{\pi r^3 p(r) + m(r)}{r^2 \left(1 - \frac{2m(r)}{r} \right)} \left(\epsilon(r) + p(r)\right)
\end{cases}
\label{eq:tov}
\end{equation}
Eqs.~\eqref{eq:tov} are solved with initial conditions for mass and pressure: $m(r=0\hbox{ km})=0$ and $p(r=0\hbox{ km})=p_c$. By varying the value for $p_c$, one generates a family of masses and radii.

The pulsation equation is defined as
\begin{equation}
\frac{dy}{dr}=  -\frac{y(r)^2}{r} -\frac{y(r)-6}{r-2m(r)} - rQ(r) \, ,
\end{equation}
where
\begin{equation}\begin{split}
\
Q(r) = \frac{4 \pi r}{r - 2m(r)} \Biggl[ [5-y(r)]\epsilon(r) + [9+y(r)]p(r) \\
+\frac{\epsilon(r)+p(r)}{\partial p(r) / \partial \epsilon(r)} \Biggr] -4 \Bigg[\frac{m(r)+4\pi^{3}p(r)}{r(r-2m(r))}\Biggr]^2 \, .
\end{split}\end{equation}

Defining $y_R$ as
\begin{equation}
y_R= y(R)-4\pi R^3\epsilon_s/M \, ,
\end{equation}
where $\epsilon_s$ is the surface energy density $\epsilon_s=\epsilon(r=R)$. For NS, $\epsilon_s=0$.

The dimensionless tidal deformability is defined as~\cite{Moustakidis:2017}
\begin{equation}
\Lambda = \frac{2}{3}k_2 \beta^{-5}
\label{eq:lambda}
\end{equation}
where $\beta = M/R$, with total mass $M$ and radius $R$, while $k_2$ is the tidal Love number, given by
\begin{equation}
k_2=\frac{8}{5}\frac{\beta^5 z}{F}
\end{equation}
with
\begin{equation}
z = (1-2\beta)^2[2-y_R+2\beta(y_R-1)] \, .
\end{equation}
and
\begin{equation}
\begin{split}
F = 6\beta (2-y_R) + 6 \beta ^2 (5y_R-8)+4\beta^3(13-11y_R) \\
+4\beta^4(3y_R-2)+8\beta^5(1+y_R)+3z \ln{1-2\beta} \, .
\end{split}
\end{equation}

\section{Parameter space exploration for RMF models}

In the literature, RMF models are often associated with a limited number of parameter sets. In the present study, we instead explore a large number of parameter sets initially provided by a Markov chain Monte Carlo (MCMC) parameter space exploration. We then filter this large number of models against several constraints, which we detail in the following subsections.

\subsection{Markov chain Monte Carlo (MCMC)}

\begin{figure}[htbp]
\centering    
\begin{subfigure}[b]{0.45\textwidth}
\centering
\includegraphics[width=\textwidth]{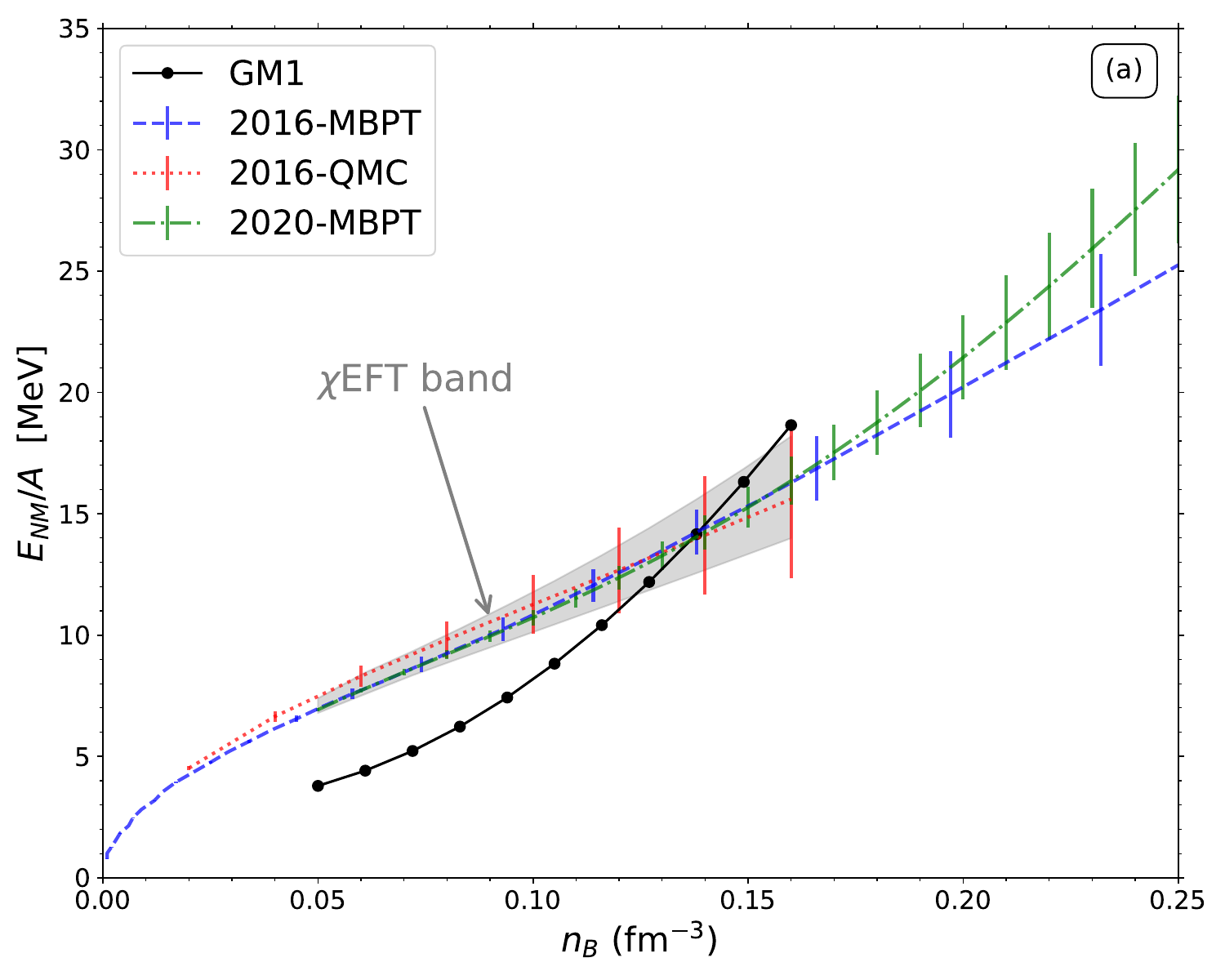} %
\end{subfigure}
\hfill
\begin{subfigure}[b]{0.45\textwidth}
\centering
\includegraphics[width=\textwidth]{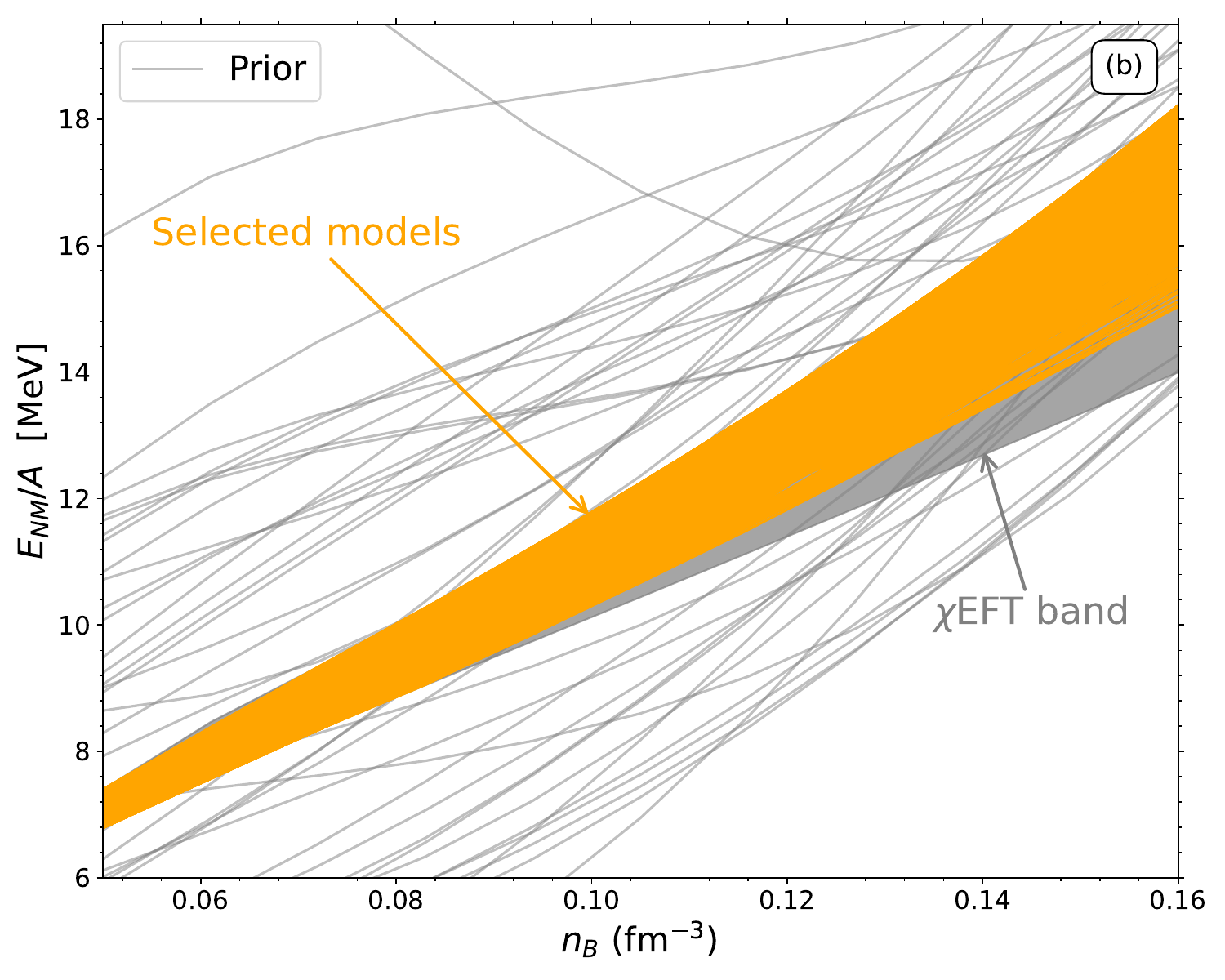}%
\end{subfigure}
\caption{Panel (a): comparison of different microscopic prediction for the binding energy in NM based on $\chi$EFT nuclear interaction~\cite{Tews:2016,Drischler:2016,Drischler:2020xEFT} with the prediction from GM1. The grey area represents an average over the several microscopic predictions, see text for more details. Panel (b): comparison of the energy per particle in NM for some of our priors (grey lines), the $\chi$EFT band identical to the one shown on the top panel (grey area), and predictions from the models selected from the NEPs given in Tab.~\ref{tab:nep_corrected_version} and $\chi$EFT band (orange area).}
\label{fig:NM_binding_energy}
\end{figure}

We employ the \texttt{emcee} python library to perform the MCMC exploration of the parameter space. The priors on the model parameters are expressed in terms of the NEPs given in Tab.~\ref{tab:nep_corrected_version}. If the set of parameters predicts at least one of the NEP to be out of the range given in Tab.~\ref{tab:nep_corrected_version}, then the prior probability is $p_\mathrm{prior}=0$. In the metropolis algorithm at the heart of the MCMC approach, a probability is assigned to each model. In our case, this probability is defined in the following way:
\begin{equation}
\log p_\mathrm{MCMC} = -\frac 1 2 \sum_{i=1}^N
\left( \frac{E(\hbox{model},n_i)-E(\hbox{$\chi$EFT},n_i)}{\Delta E(\hbox{$\chi$EFT},n_i)}\right)
\end{equation}
where $i$ runs over the densities $n_i$, and where $E(\hbox{$\chi$EFT},n_i)$ and $\Delta E(\hbox{$\chi$EFT},n_i)$ represent the centroids and uncertainties of the $\chi$EFT band defined hereafter, see Fig.~\ref{fig:NM_binding_energy}(a). In case the RMF model is inside the $\chi$EFT band, the probability becomes constant: if $\log p_\mathrm{MCMC} \ge -N/2$, then $\log p_\mathrm{MCMC} = -N/2$.

The MCMC is initialized with a set of 50 models compatible with Tab.~\ref{tab:nep_corrected_version} and the MCMC method is evolved until the total number of different and viable models reaches 10,000.

\begin{table}[t]
\centering
\tabcolsep=0.2cm
\def\arraystretch{1.4}
\begin{tabular}{ccccc}
\hline
&  \multicolumn{2}{c}{$\chi$EFT band} & \multicolumn{2}{c}{$\chi$EFT+NEP band} \\
Densities & $E/A_{\min}$ & $E/A_{\max}$ & $E/A_{\min}$ & $E/A_{\max}$ \\
fm$^{-3}$ & MeV & MeV & MeV & MeV \\
\hline
0.05 & 6.79 & 7.40 & 6.79 & 7.40 \\
0.06 & 7.57 & 8.46 & 7.57 & 8.40 \\
0.07 & 8.34 & 9.34 & 8.34 & 9.34 \\
0.08 & 9.05 & 10.27 & 9.04 & 10.27 \\
0.09 & 9.74 & 11.23 & 9.85 & 11.23 \\
0.11 & 10.44 & 12.24 & 10.67 & 12.24 \\
0.12 & 11.14 & 13.29 & 11.50 & 13.29 \\
0.13 & 11.85 & 14.41 & 12.38 & 14.41 \\
0.14 & 12.55 & 15.60 & 13.25 & 15.60 \\
0.15 & 13.27 & 16.86 & 14.12 & 16.86 \\
0.16 & 13.99 & 18.19 & 15.03 & 18.19 \\
\hline
\end{tabular}%
\caption{Boundary for the band extracted from the microscopic predictions in NM based on $\chi$EFT nuclear interactions~\cite{Tews:2016,Drischler:2016,Drischler:2020xEFT}. See text for more details.}
\label{tab:band}
\end{table}
Predictions for the energy per particle in NM based on $\chi$EFT nuclear interactions~\cite{Tews:2016,Drischler:2016,Drischler:2020xEFT} are shown in Fig.~\ref{fig:NM_binding_energy}(a). The uncertainty quantifications (UQs) performed by the authors of these calculations are represented by vertical bars. These UQs are performed differently by these authors implying some complexities in their interpretation. We, therefore, consider an agnostic approach averaging over these microscopic predictions: at each density in the region $n\in[0.05,0.16]$~fm$^{-3}$, each of these $\chi$EFT predictions are represented by a Gaussian distribution centered at their centroid with a width equal to their UQ. A global distribution is then obtained by adding them, and the band (grey area) represents the boundaries of the global distribution defined as its centroid $\pm$ its standard deviation. The values we obtain are given in Tab.~\ref{tab:band}.

As a reference, the prediction for NM energy per particle from GM1 Lagrangian \cite{PhysRevLett.67.2414} is also shown. We conclude that the Lagrangian GM1 is not in good agreement with the $\chi$EFT band.

We show in Fig.~\ref{fig:NM_binding_energy}(b) a few examples for the priors model that do not agree with the $\chi$EFT band (grey lines), the $\chi$EFT band (grey area), and the selected models (orange area). There is a wide overlap between the selected models and the $\chi$EFT band, except for the low values of the energy per particle around saturation density. The reason is due to the interval for $E_\sat$ and $E_{\sym,2}$ considered in Tab.~\ref{tab:nep_corrected_version}. Since we have $E_\NM/A \approx E_\sat+E_{\sym,2}$, the lower values of $E_\NM/A$ around saturation density reflect the choice for the range of $E_\sat$ and $E_{\sym,2}$ given in Tab.~\ref{tab:nep_corrected_version}.
To populate the lower values of $E_\NM/A$ around saturation density we shall explore values for $E_\sat$ and/or $E_{\sym,2}$ which are lower than the ones suggested in Tab.~\ref{tab:nep_corrected_version}. These lower values are however excluded by the phenomenology of finite nuclei. So the double constraints imposed to the selected models to satisfy the NEP range given in Tab.~\ref{tab:nep_corrected_version} and the $\chi$EFT band, result in a band which is shown in orange in Fig.~\ref{fig:NM_binding_energy}(b). The boundaries for this band, allowed by $\chi$EFT and NEPs from Tab.~\ref{tab:nep_corrected_version}, are given in Tab.~\ref{tab:band}.

\begin{figure}
\centering
\includegraphics[width=1.0\linewidth]
{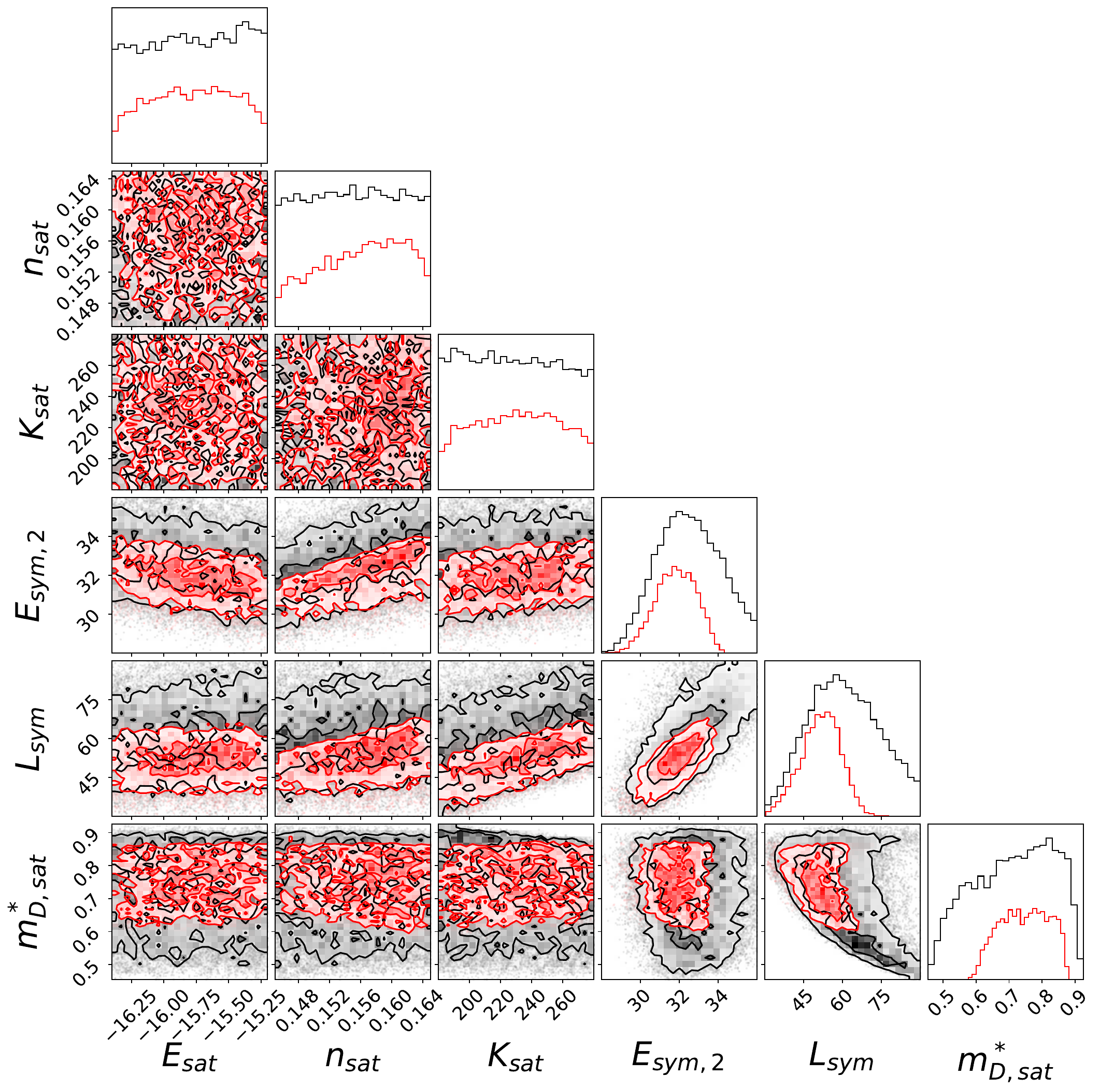}
\caption{Corner plot comparing the distributions and correlations between NEPs for the models selected from the NEPs in Tab.~\ref{tab:nep_corrected_version} and the $\chi$EFT band (black) and for the models that are further selected on the tidal deformability from GW170817 (red). See text for more details.}
\label{fig:corner_plot_tidal}
\end{figure}

Considering the NEPs given in Tab.~\ref{tab:nep_corrected_version} and the $\chi$EFT band shown in Fig.~\ref{fig:NM_binding_energy}, the distribution of NEP and their mutual correlations are shown in the corner plot in Fig.~\ref{fig:corner_plot_tidal} (black). While the isoscalar NEPs ($E_\sat$, $n_\sat$, and $K_\sat$) are rather flat, the isovector NEPs ($E_{\sym,2}$ and $L_\sym$) are peaked at $E_{\sym,2}=32\pm 2$~MeV and $L_\sym=60\pm10$~MeV. This is a consequence of the selection against the $\chi$EFT band, which is defined for NM. The distribution for the Dirac mass $m^*_{D,\sat}$ is shifted towards the largest values in the range defined in Tab.~\ref{tab:nep_corrected_version}.

In the following, we filter these initial 10,000 models employing astrophysical data, as described hereafter.

\subsection{Astrophysical data}

We now detail the constraints imposed by several astrophysical observations on our model selection. A preliminary remark is, however, necessary here. In the comparison of our models to astrophysical data, we assume a nucleonic type of EOS describing the source of the astrophysical data. Since this type of EOS is nucleonic, we do not explore the possibility of phase transitions in the core. Therefore, while an exclusion of all our selected EOS would be a sign that they are not able to describe astrophysical data (which do not occur), the fact that we find a set of models surviving the astrophysical selection cannot be interpreted as a validation of the nucleonic hypothesis. We can only conclude that nucleonic EOSs are not ruled out by astrophysical data. 

To produce the EOSs for NSs, we impose beta-equilibrium in uniform matter and we connect to the BPS EOS for the crust, as explained before.

\subsubsection{Tidal deformability from GW170817}

For a given EOS and a fixed mass $M$, the dimensionless tidal deformability is obtained from Eq.~\eqref{eq:lambda}. For GW170817, we select all the models for which the tidal deformability is $\Lambda\in[70,770]$ for a NS with a mass $M=1.4M_\odot$.

To show the impact of GW170817 on our model selection, the posterior distributions for the NEPs given in Tab.~\ref{tab:nep_corrected_version} are shown in Fig.~\ref{fig:corner_plot_tidal} (red). The difference between the black and the red lines represents the impact of GW170817 on the model selection. The most noticeable impact of the constraints from $\chi$EFT is observed on the isovector NEP $E_{\sym,2}$ and $L_\sym$ for which the distributions are shifted towards lower values compared to the prior ones. 
After the selection from $\chi$EFT, we obtain $E_{\sym,2}\le 34$~MeV and $L_\sym\le 70$~MeV. The Dirac effective mass is also impacted since the new distribution excludes both the lower and higher values indicated in Tab.~\ref{tab:nep_corrected_version}. We obtain, after the selection from $\chi$EFT, that $0.6\le m^*_{D,\sat}<0.9$. In more details, the parameters $E_{\sym,2}$ and $L_\sym$ peak at $32$~MeV and $52$~MeV, respectively and $m^*_{D,\sat}$, shows a slight peak at $0.78$.  

Since the probabilities are not normalized, the difference between the priors (in black) and the selected models (in red) reflects the effectiveness of GW170817: from 10,000 initial models, we obtain about 7,018 after the selection against the tidal deformability from GW170817.

\subsubsection{Maximum mass}

\begin{table}[t]
\centering
\renewcommand{\arraystretch}{1.5} 
\setlength{\tabcolsep}{10pt}     
\begin{tabular}{cccc}
\hline
\textbf{Sources} & $i$ & $M_{\obs,i}$ & Ref. \\
 & & [$M_{\odot}$] & \\
\hline
PSR J1614–2230 &   & $1.970\pm 0.040$ & \cite{Demorest:2010} \\
&   & $1.928\pm 0.017$ & \cite{Fonseca:2016} \\
& 1 & $1.908\pm 0.016$ & \cite{Arzoumanian:2018} \\
&   & $1.922\pm 0.015$ & \cite{Alam:2021} \\
&   & $1.937\pm 0.014$ & \cite{Agazie:2023} \\
PSR J0348+0432 & 2 & $2.01\pm 0.04$ & \cite{Antoniadis:2013} \\
MSP J0740+6620 &   & $2.14\pm 0.10$ & \cite{Cromartie:2019}\\
& 3 & $2.08\pm 0.07$ & \cite{Fonseca:2021} \\
&   & $1.99\pm 0.07$ & \cite{Agazie:2023} \\
PSR J2215+5135 & 4 & $2.27\pm 0.15$ & \cite{Linares:2018}\\
\hline
\end{tabular}
\caption{List of recently observed masses ($M_{\obs,i}$) among which we select four constraints, identified with an integer value for $i$. For two of these pulsars, there are several measurements of the masses, and we select only one of them.}
\label{tab:model_pulsar}
\end{table}

Recent observations of massive NS from radio astronomy are listed in Tab.~\ref{tab:model_pulsar}. Two of these four pulsars have been measured several times, either by the same team or by different teams, and the latest measurement is not always the best to consider. In Ref.~\cite{Agazie:2023} for instance, the authors do not regard their measurement, as superseding the previous one from Ref.~\cite{Fonseca:2021}. They are even satisfied by the fact that their result is consistent with the one from Ref.~\cite{Fonseca:2021} within 1$\sigma$. For this reason, we select the measurement from Ref.~\cite{Fonseca:2021} to constrain the TOV mass. The final result does not, however, depend substantially on our choices.

\begin{figure}[t]
\centering
\includegraphics[scale=0.35]{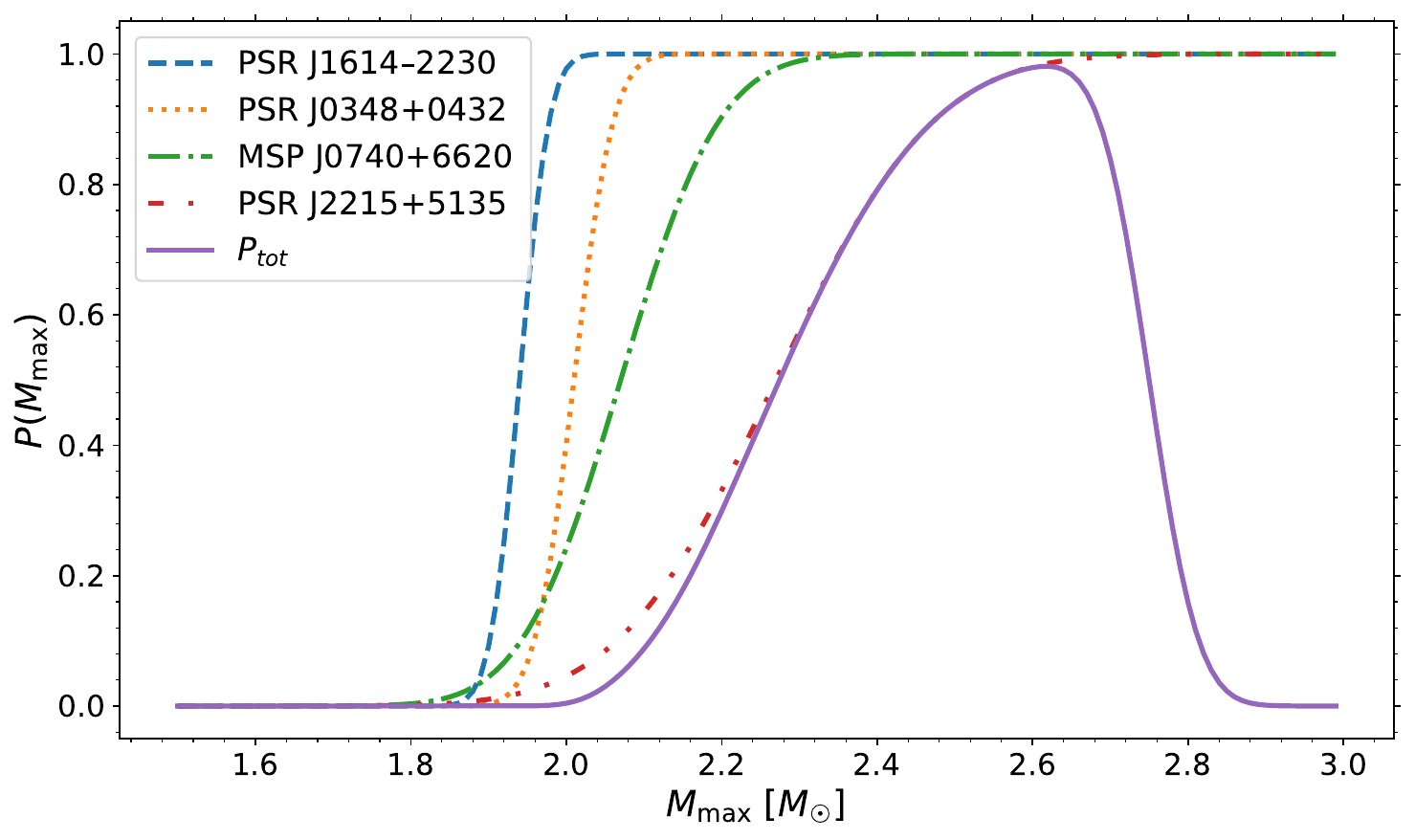}
\caption{The probabilities $P(M_\max)$ associated with the four observations are given in dashed, dotted, and dash-dotted lines. The solid line refers to the total probability $P_{\text{tot}}(M_{\mathrm{max}})$.}
\label{fig:tov_mass}
\end{figure}

Each of these observations of massive NSs produces a lower estimate for the TOV mass. Considering the uncertainty reported in these measurements, the probability distributions for the TOV mass are plotted in Fig.~\ref{fig:tov_mass} (dashed, dotted and dash-dotted lines). These probabilities are defined as
\begin{equation}
P_i(M_\max) = \frac{\text{erf}[f_i(M_\max)]+1}{2}, 
\label{eq:mass_proba}
\end{equation}
where $\text{erf}(z)$ is the error function, $i$ indicates the observed data reported in Tab.~\ref{tab:model_pulsar}, and the function $f_i(M)$ is defined as
\begin{equation}
f_i(M) = \frac{M-M_{\obs,i}}{\sqrt{2} \, \sigma_{M,i}},
\end{equation}
where $M_\obs$ and $\sigma_M$ are the observed mass and the associated error, respectively.
The probability~\eqref{eq:mass_proba} is assigned to a given EOS with the TOV mass $M_\max$ expressed in solar mass units.

Since the two NSs that have produced GW170817 are believed to have produced a black hole~\cite{Abbott:2017}, the sum of their masses defined an upper boundary for the TOV mass. We introduce the associated probability $P_\mathrm{GW170817}(M_\max)$ defined as $P_\mathrm{GW170817}(M_\max)= (1-\text{erf}[f_5(M_\max)])/2$, with $M_{\obs,5}=2.74$~M$_\odot$ and $\sigma_{M,5}=0.04$~M$_\odot$ obtained from the waveform analysis for low-spin priors~\cite{Abbott:2017}.

Finally, since all these constraints are independent of each other, one can define a total probability by multiplying them together. We obtain
\begin{equation}
P_{\text{tot}}(M_\max) = P_\mathrm{GW170817}(M_\max) \prod_{i=1}^4 P_i(M_\max) \, ,
\end{equation}
which is represented in Fig.~\ref{fig:tov_mass} (solid line). 
Note that the lower boundary of the distribution is strongly dependent on the mass observation of PSR J2215+5135, which remains quite uncertain. We, however, decided to keep this observation to define the TOV mass probability distribution to select stiff models at high density (or high mass). It is indeed difficult to have models that change from soft for canonical mass NS to stiff at higher densities (thus for massive NSs). Anticipating our final results, we show that the RMF approach presented here is a good candidate for obtaining such behavior.

\begin{figure}[t]
\centering
\includegraphics[width=1\linewidth]{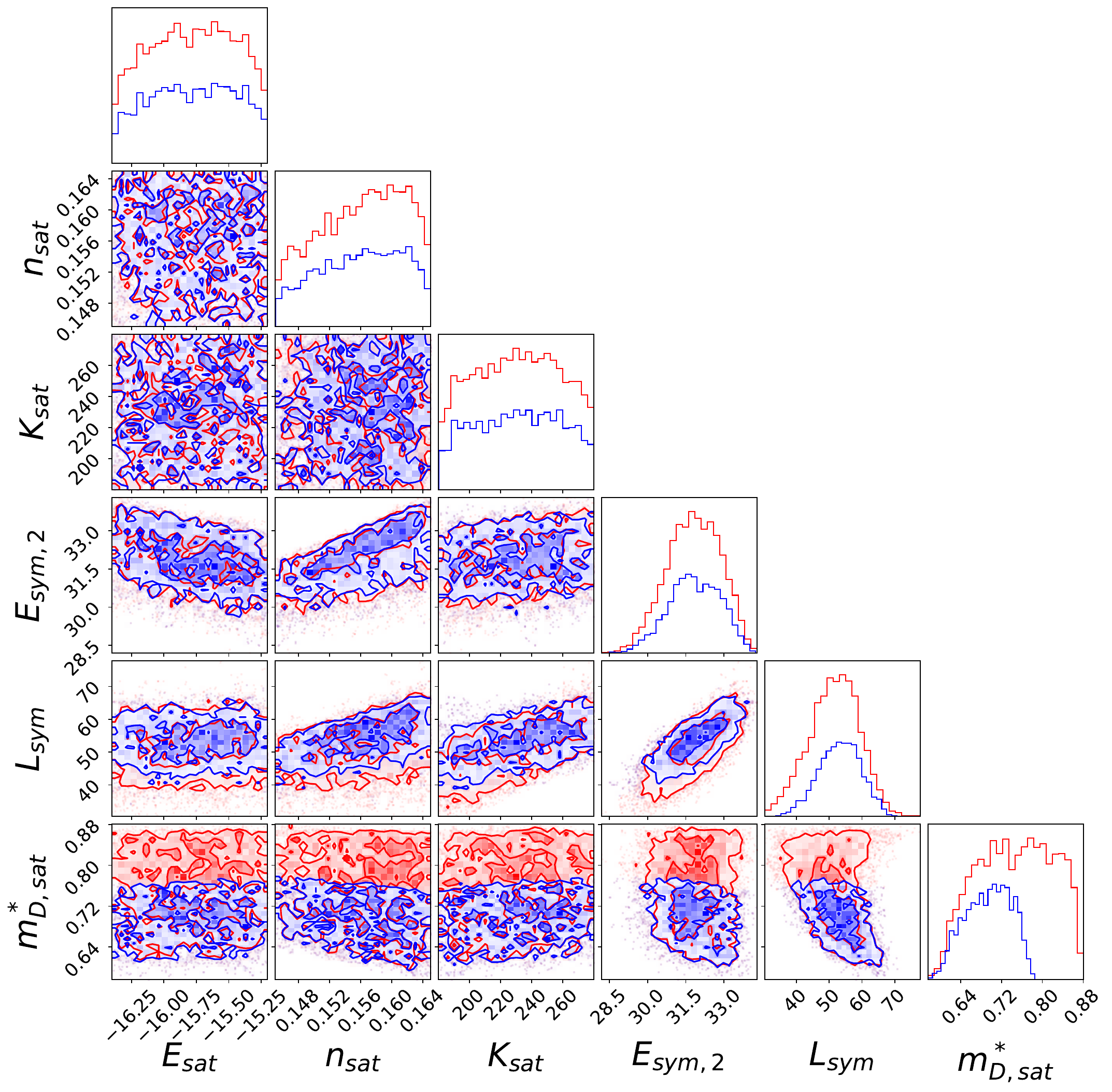}
\caption{Same as Fig.~\ref{fig:corner_plot_tidal} for models satisfying the constraint from the TOV mass (blue) on top of the previous ones (red). Note that the red distribution is identical to the one in Fig.~\ref{fig:corner_plot_tidal}.}
\label{fig:comparison_plot_2_3}
\end{figure}

We show in Fig.~\ref{fig:comparison_plot_2_3} a comparison of the distribution of parameters after the maximum mass selection (blue), for which all models predicting a TOV mass lower than $2M_\odot$ are removed, see Fig.~\ref{fig:tov_mass}. As a reference, the result of the MCMC parameter exploration and the constraint from GW170817 is also shown (red). The result of this further selection is to isolate 3945 models (blue) out of the 7018 ones previously selected (red). Since the constraint on the maximum mass is applied to beta-equilibrium matter in NS, which is quite neutron-rich, the result of the selection impacts more the isovector channel, $E_{\sym,2}$ and $L_\sym$, than the isoscalar one. Interestingly, the Dirac mass is also impacted and now peaks at about 0.70.

\begin{figure}[t]
\centering    \includegraphics[width=1\linewidth]{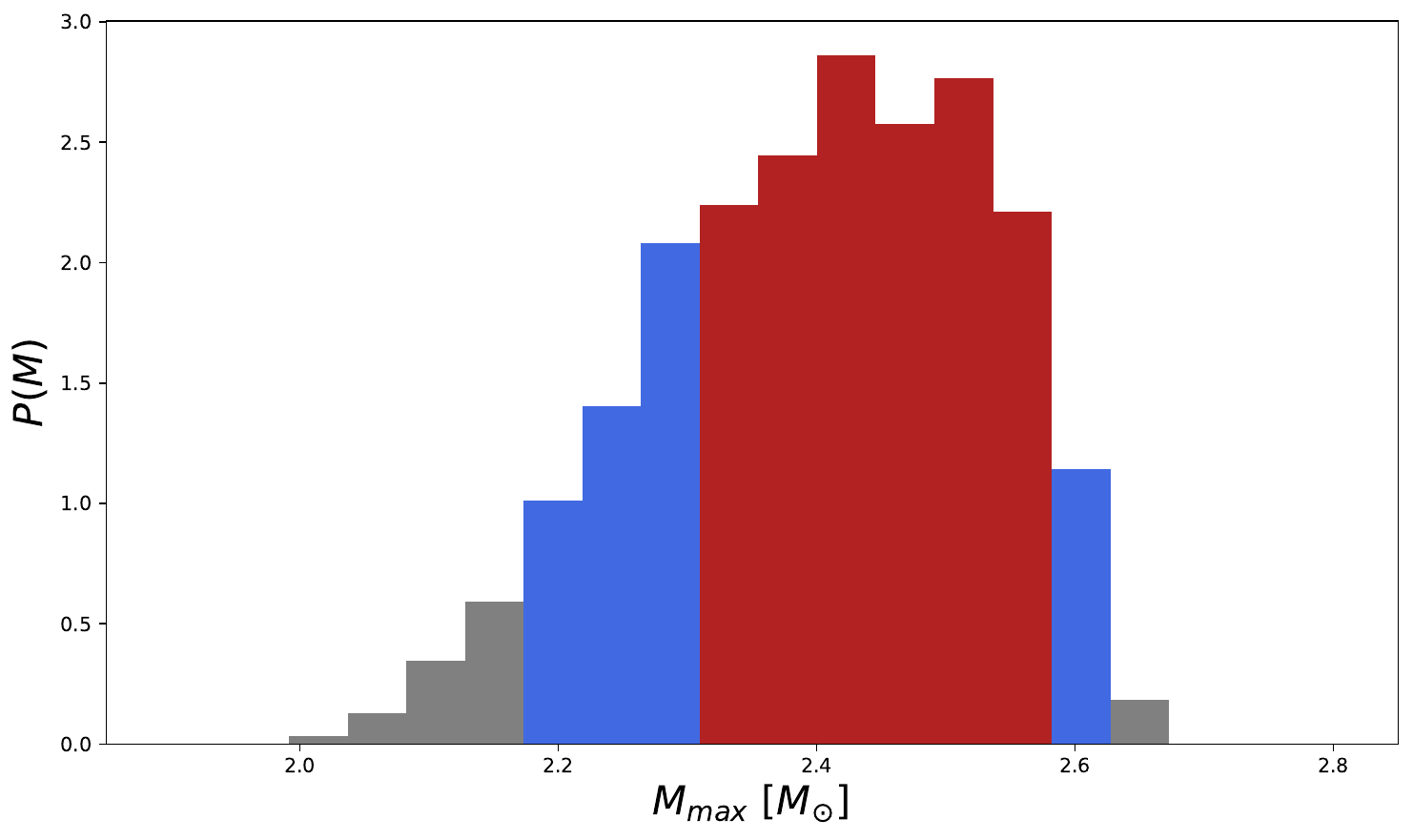}
\caption{Histogram showing the normalized probability as a function of the maximum mass. The models within the 68\% probability range are indicated in red, those corresponding to the 95\% probability range in blue, and the remaining models in grey.}
\label{fig:histo}
\end{figure}

In Fig.~\ref{fig:histo} we represent the distribution of models as a function of the TOV mass $M_\max$. The model distribution corresponds to our final selection, i.e, selecting according to NEPs, $\chi$EFT band, GW170817 tidal deformability and the distribution of maximum mass. In red are shown the models representing 68\% of the total probability, in blue the complement to reach 95\% of the total probability, and the rest is shown in grey.

\subsubsection{Comparison to NICER analyses}
\label{sec:astro:nicer}

\begin{figure}[t]
\centering
\includegraphics[width=1\linewidth]{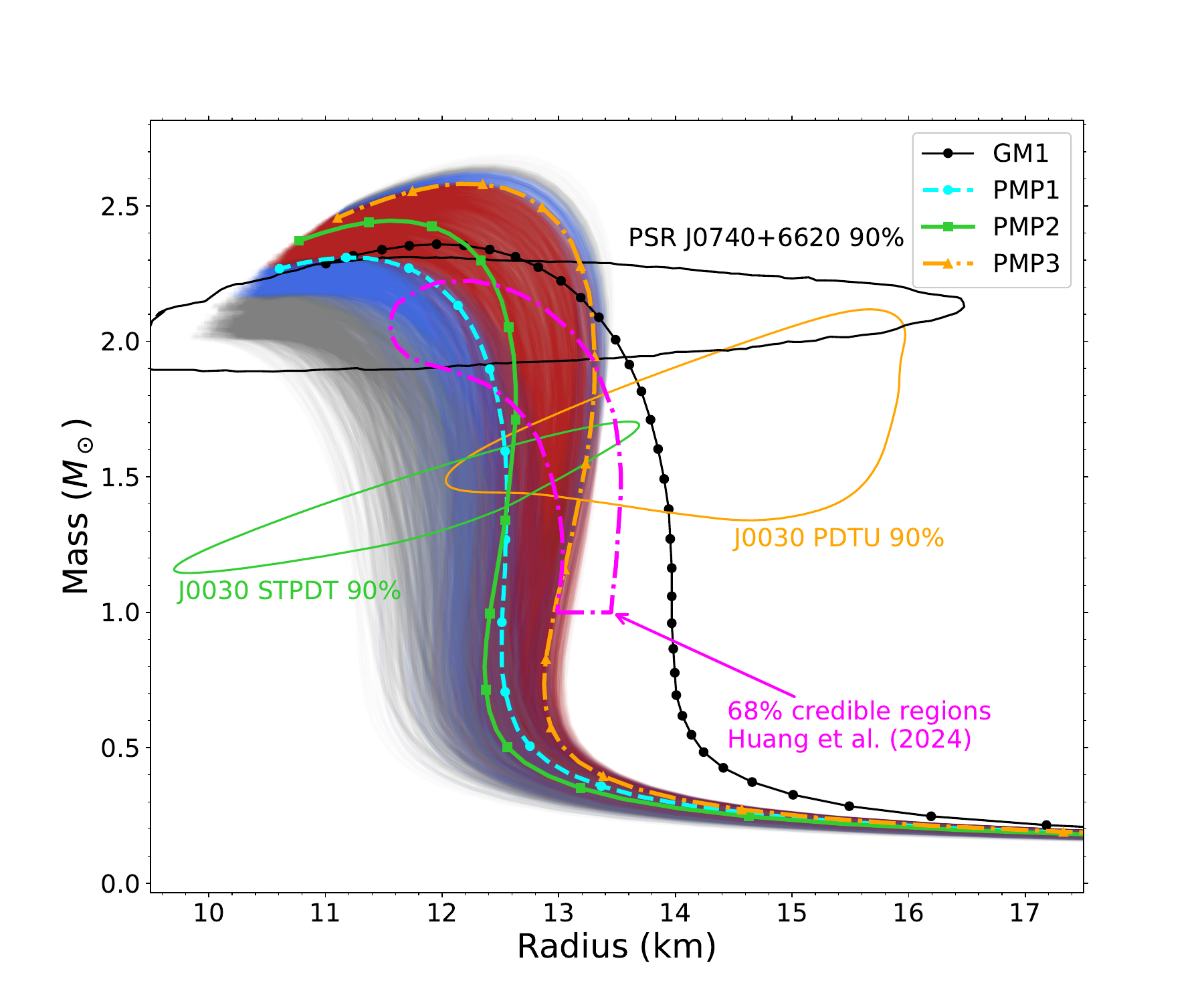}
\caption{Mass-radius relation solution of the TOV equations for each finally selected EOS. The colors (red, blue, grey) correspond to the distribution shown in Fig.~\ref{fig:histo}. The pink dash-dotted line is referred to the 68\% credible regions of Fig. 7 in \cite{Huang:2023grj}.}
\label{fig:RM_plot}
\end{figure}

\begin{figure}[t]
\centering
\includegraphics[width=1\linewidth]{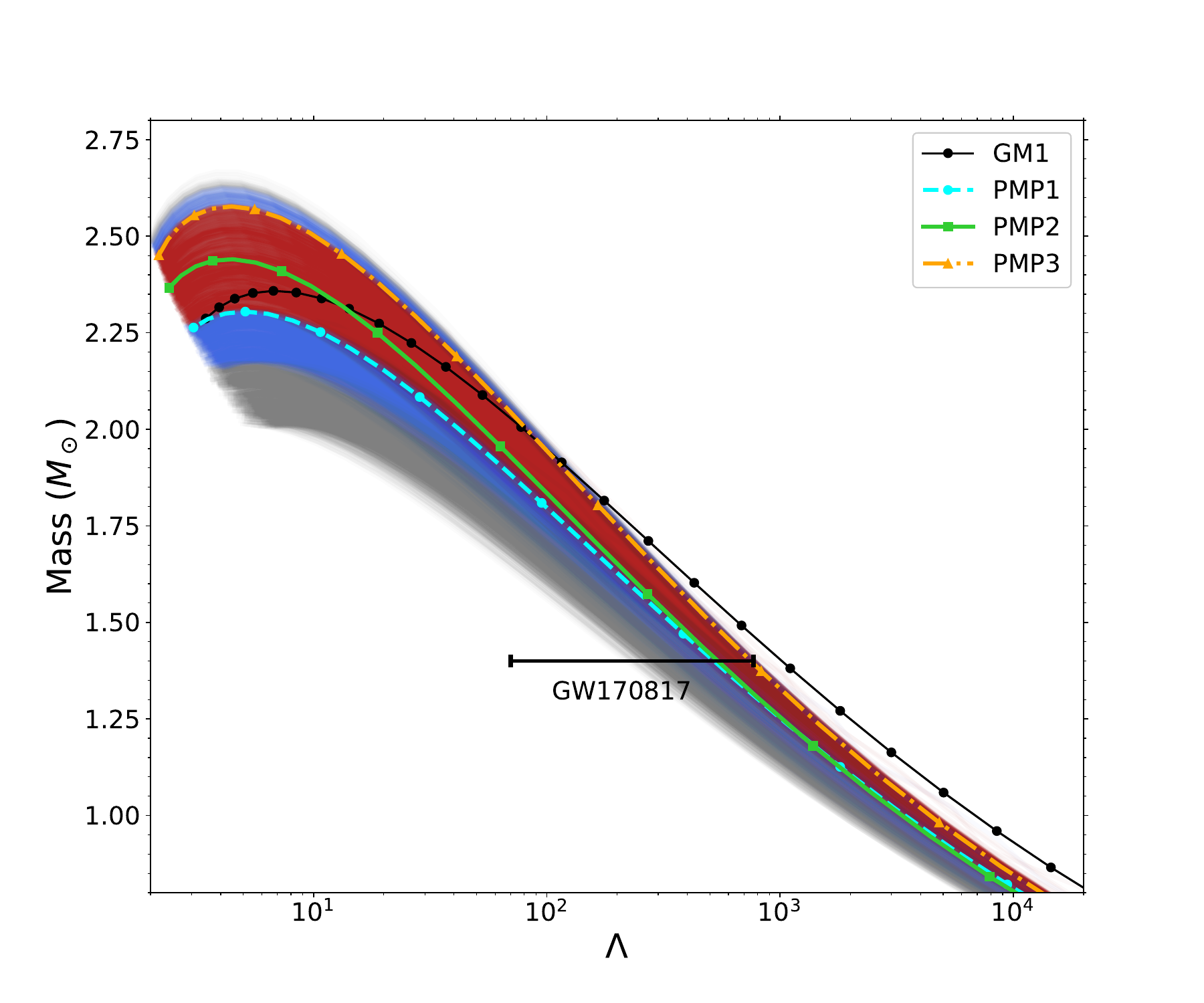}
\caption{Mass-$\Lambda$ relation solution of the TOV equations for each finally selected EOS. The colors (red, blue, grey) correspond to the distribution shown in Fig.~\ref{fig:histo}.}
\label{fig:tidal_plot}
\end{figure}

The mass-radius relations for all the finally selected models are shown in Fig.~\ref{fig:RM_plot}.
The contours correspond to the 90\% confidence interval and are associated with the observation of PSR J0030+0451 \cite{Vinciguerra_2024} and PSR J0740+6620 \cite{Miller_2021}\ by NICER team. In the case of PSR J0030+0451 two contours are suggested recently depending on the hot-spot model (PDTU and STPDT), with a small overlap between them. Since the authors, among the two observational data analysis, have not chosen which is preferable, we represent them both.
Let us remark, however, that our selected best models (red) overlap with the overlap between the two contours for PSR J0030. There is also a very good overlap between our finally selected models and the 90\% confidence interval for PSR J0740+6620. We therefore conclude that our finally selected models are in good agreement with the present analyses by NICER.

The maximum mass of each model exceeds 2.1$M_\odot$, with no model exhibiting a maximum mass above 2.7$M_\odot$. This finding is consistent with the assumed probability function, $P(M_\max)$. 

\begin{table*}[tb]
\centering
\tabcolsep=0.18cm  
\def\arraystretch{1.3}  
\begin{tabular}{lccccccccccccc}
\hline
\textbf{Models} & $E_{\sat}$ & $n_{\sat}$ & $K_{\sat}$ & $E_{\sym,2}$ & $L_{\sym}$ & $m^*_{D,\sat}$ & $M_\max$ & $g_{\sigma}/m_\sigma$ & $g_{\omega}/m_\omega$ & $g_{\rho}/m_\rho$ & $a_{\rho}$  & $b$ & $c$ \\
 & (MeV) & (fm$^{-3}$) & (MeV) & (MeV) & (MeV) &  & ($\msun$) & (fm) & (fm) & (fm) &  & $\times 100$ & $\times 100$ \\
\hline
PMP1 & $-15.40$ & $0.159$ & $234.9$ & $31.71$ & $56.6$ & $0.687$ & $2.31$ & 3.472 & 2.686 & 1.966 & 0.405 & 0.413 & -0.455 \\
PMP2 & $-15.41$ & $0.154$ & $187.9$ & $31.36$ & $49.0$ & $0.662$ & $2.45$ & 3.676 & 2.862 & 1.962 & 0.527 & 0.414 & -0.552 \\
PMP3 & $-15.66$ & $0.152$ & $217.1$ & $31.93$ & $57.1$ & $0.631$ & $2.58$ & 3.803 & 3.035 & 1.976 & 0.488 & 0.302 & -0.396 \\
PMP$_\text{R2}$ & -15.43 & 0.156 & 219.12 & 32.20 & 56.59 & 0.662 & 2.43 & 3.623 & 2.842 & 1.996 & 0.444 & 0.369 & -0.459 \\
PMP$_\text{R4}$ & -16.28 & 0.152 & 207.82 & 32.43 & 57.54 & 0.629 & 2.58 & 3.836 & 3.036 & 2.006 & 0.488 & 0.327 & -0.433 \\ 
PMP$_\text{R6}$ & -15.44 & 0.165 & 197.65 & 33.28 & 64.76 & 0.624 & 2.50 & 3.697 & 2.932 & 1.921 & 0.449 & 0.341 & -0.470 \\
\hline
\end{tabular}%
\caption{Parameters for PMP1, PMP2, PMP3, PMP$_\text{R2}$, PMP$_\text{R4}$ and PMP$_\text{R6}$ RMF models. The models PMP$_\text{R2}$, PMP$_\text{R4}$ and PMP$_\text{R6}$ predict $R_{\mathrm{dURCA}}$ equal to 2, 4 and 6 km respectively for $1.4M_\odot$ NS. See text for more details.}
\label{tab:PMP1_NEP}
\end{table*}

As a reference, we also show in Fig.~\ref{fig:RM_plot} the mass-radius relation obtained for GM1. As often with RMF approach, the radius of a canonical mass NS (about $1.4M_\odot$) is around 14~km, see e.g. Ref.~\cite{Fortin:2014mya}, while our best models predict lower values for the radius, $R_{1.4M_\odot}\in[12.3,13.3]$~km.

The mass-radius predictions for PMP1-PMP3 RMF models are also shown in Fig.~\ref{fig:RM_plot}. The models PMP1 and PMP3 are localized at the boundary of the 68\% confidence interval (red region), while PMP2 is more central. It is selected from the TOV mass, which is intermediate between the two ones predicted by PMP1 and PMP3, see Tab.~\ref{tab:PMP1_NEP}.

\begin{figure}[t]
\centering
\includegraphics[width=1\linewidth]{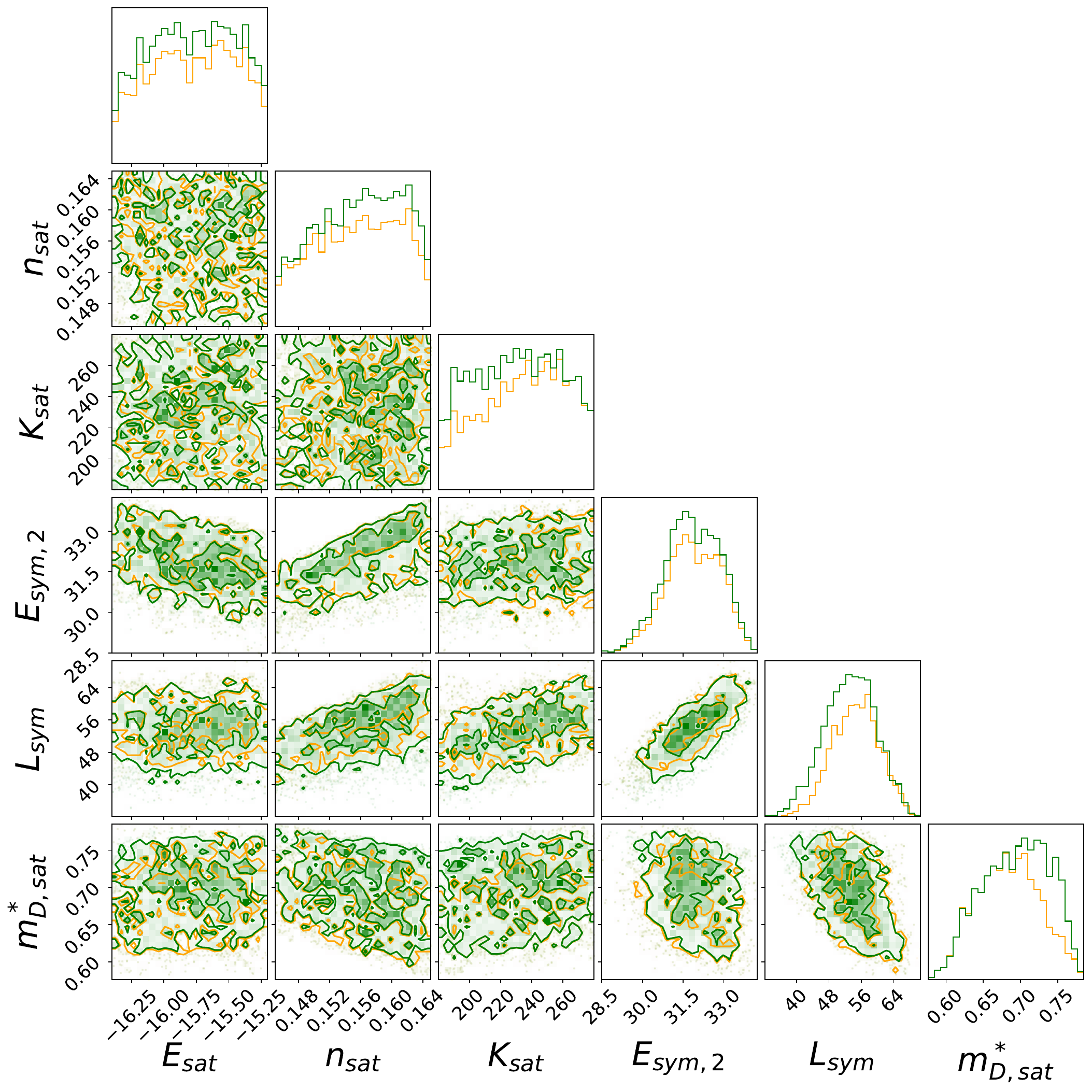}
\caption{The models that satisfy the constraints on $\chi$EFT, tidal deformability and the additional constraint given by NICER data are reported in orange for PDTU, while in green for STPDT}    \label{fig:comparison_plot_NICER}
\end{figure}

The impact of the two contours for PSR J0030 recently suggested by NICER is shown in Fig.~\ref{fig:comparison_plot_NICER}. The distribution considering the model PDTU for the hot spot (yellow) is compared to the distribution for the model STPDT (green). There is a weak impact of these two possible observations on our models, and additionally, these two distributions do not reduce substantially the number of models in our set. The main reason is that our best models have already a very good overlap with the two contours as shown in Fig.~\ref{fig:RM_plot}.

\section{Predictions based on the selected RMF models}

In the previous section, we have selected a set of about $5,000$ RMF models consistent with the expectation for NEPs, the $\chi$EFT band, the tidal deformability of GW170817 and the present knowledge on the distribution of maximum mass. In this section, we employ these models, for which we identify 68\% and 95\% confidence intervals shown in Fig.~\ref{fig:histo}, to predict properties of NSs.

We remind that we have considered that the NS's core is composed only of nucleons and leptons. Our predictions are therefore conditioned by two main hypotheses: I) the RMF approach is correct and ii) the composition of the core is nucleons and leptons only.

\subsection{Equation of state}

\begin{figure}
\centering
\includegraphics[width=1\linewidth]{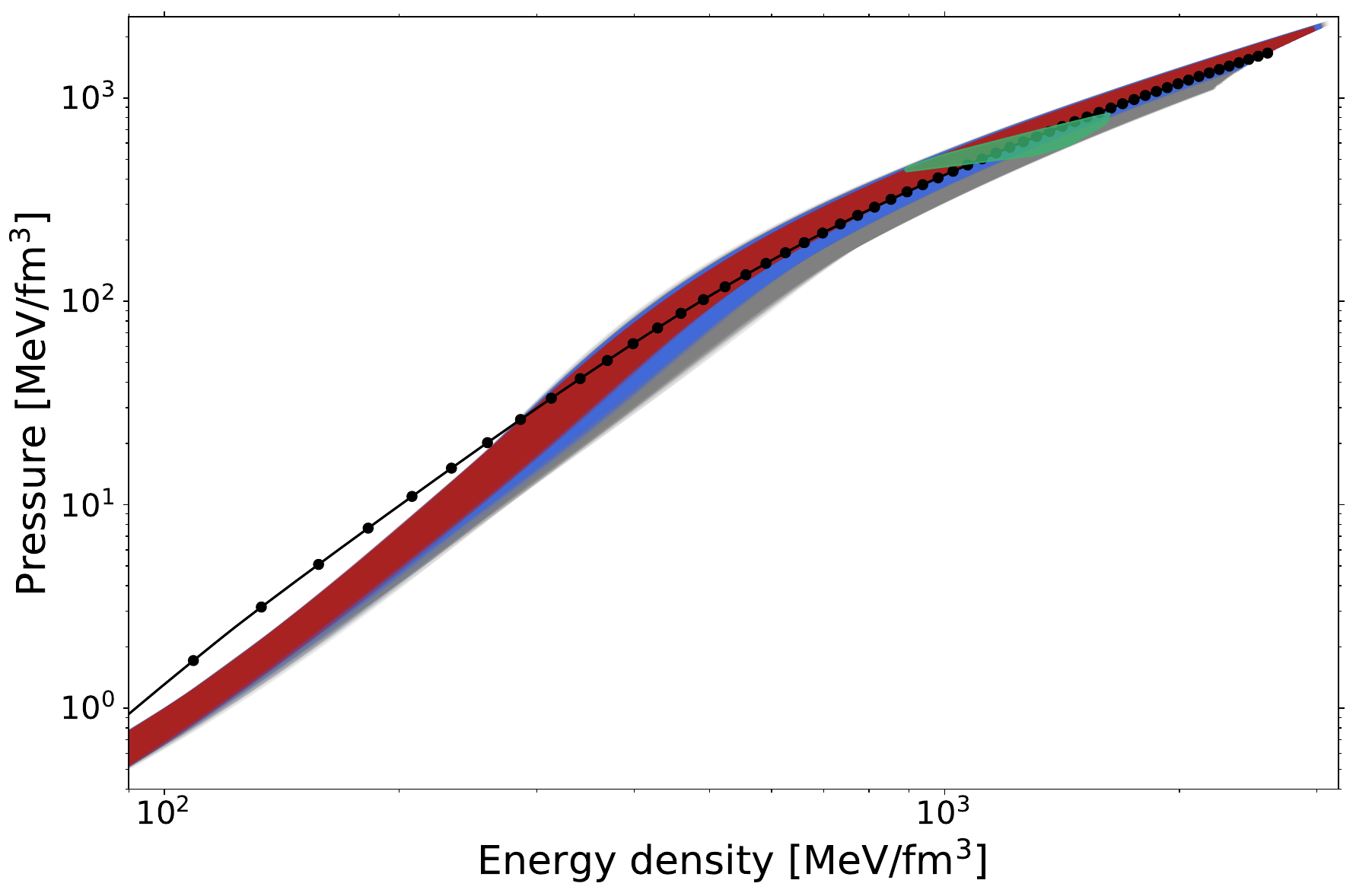}
\caption{EOS for the selected models with same color code as the one shown in Fig.~\ref{fig:histo}.
For reference, the EOS generated from GM1 RMF is also shown (black solid line). The EOS are generated for baryon densities between 0.08 and 2.0~fm$^{-3}$.
The green region corresponds to the upper boundary of the stable branch.}
\label{fig:EOS}
\end{figure}

\begin{figure}
\centering
\includegraphics[width=1\linewidth]{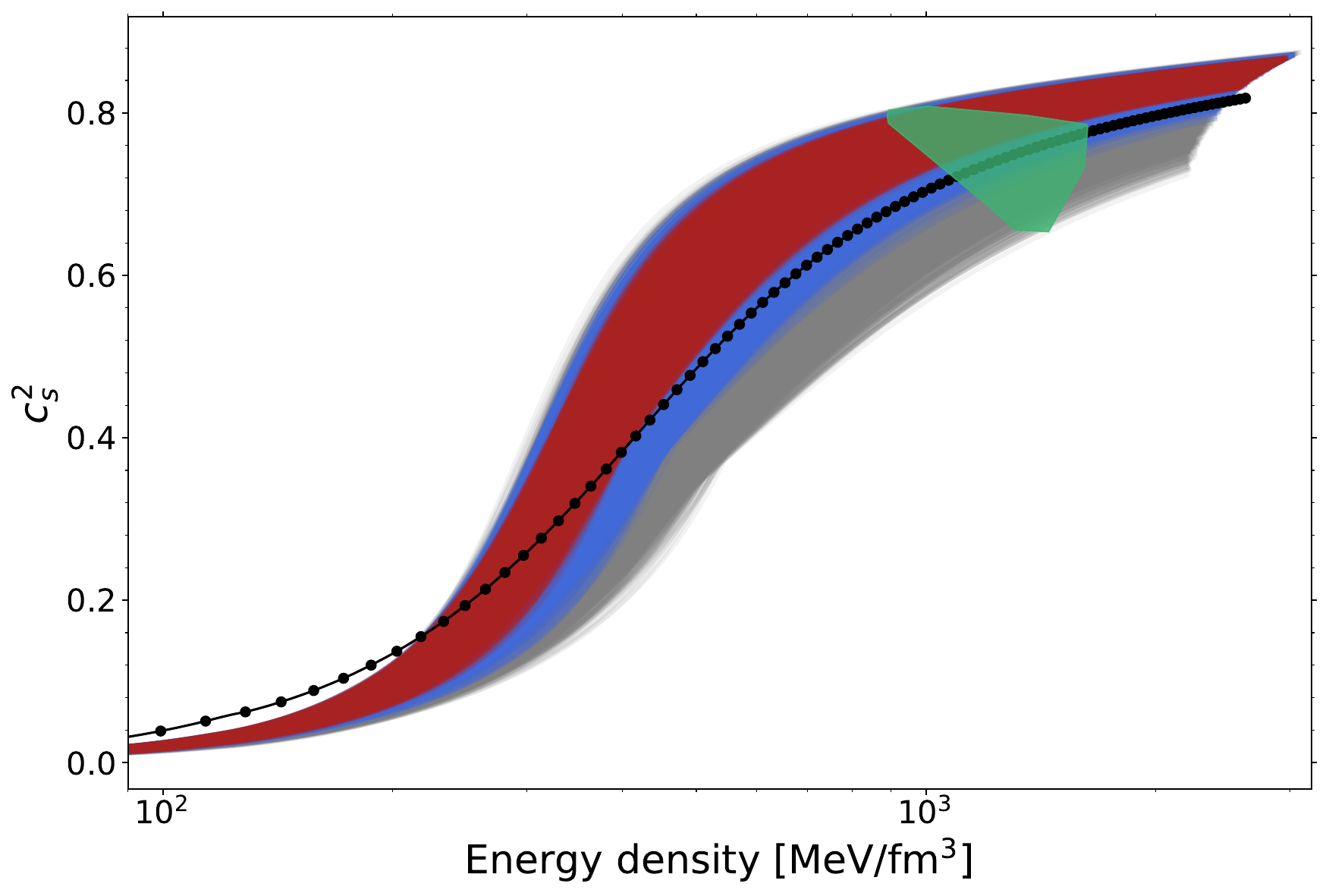}
\caption{Same as Fig.~\ref{fig:EOS} for the square of the sound speed $c_s^2(\epsilon)$.}
\label{fig:cs}
\end{figure}

The EOS $P(\epsilon)$ and the sound speed $c_s^2=d P/d\epsilon$ are shown in Figs.~\ref{fig:EOS} and \ref{fig:cs} for the finally selected models adopting the sale color code as in Fig.~\ref{fig:histo} for the 68\% and 95\% confidence intervals. As a reference, the prediction for the EOS based on GM1 EOS is also shown in the figures. It is interesting to remark that for most of the energy-densities shown in the figure, GM1 is softer than the 68\% of the selected models, while it predicts a larger radius up to about $2M_\odot$, see Fig.~\ref{fig:RM_plot}. The reason is that the range of energy densities that matter for the GM1 mass-radius relation is located below 400~MeV~fm$^{-3}$. For this range of energy-densities, GM1 is more repulsive than most of our selected models, as shown in Fig.~\ref{fig:cs}.

\subsection{Central densities}

\begin{figure}[t]
\centering
\begin{subfigure}[b]{0.45\textwidth}
\centering
\includegraphics[width=\textwidth]{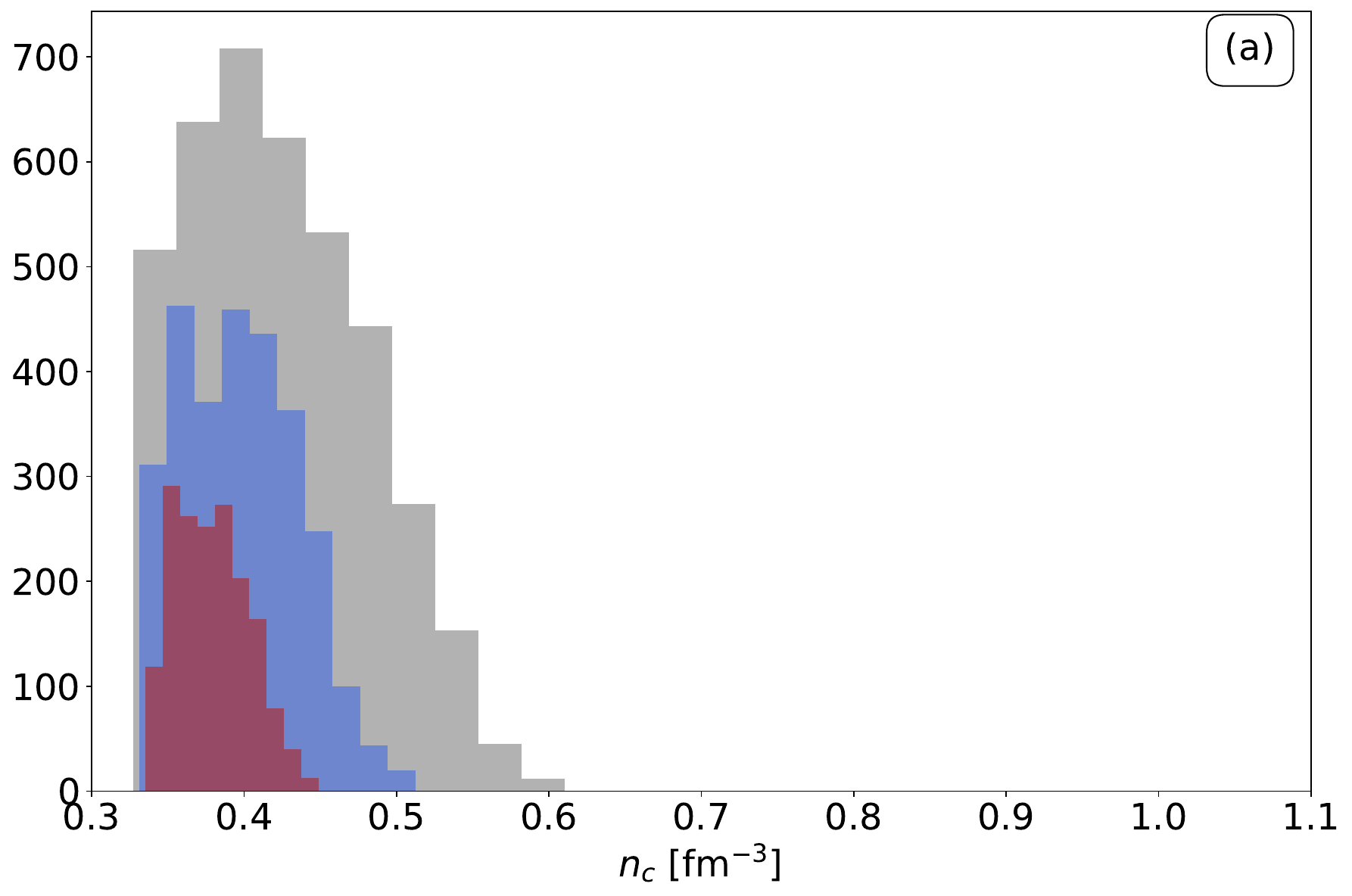} 
\end{subfigure}
\hfill
\begin{subfigure}[b]{0.45\textwidth}
\centering
\includegraphics[width=\textwidth]{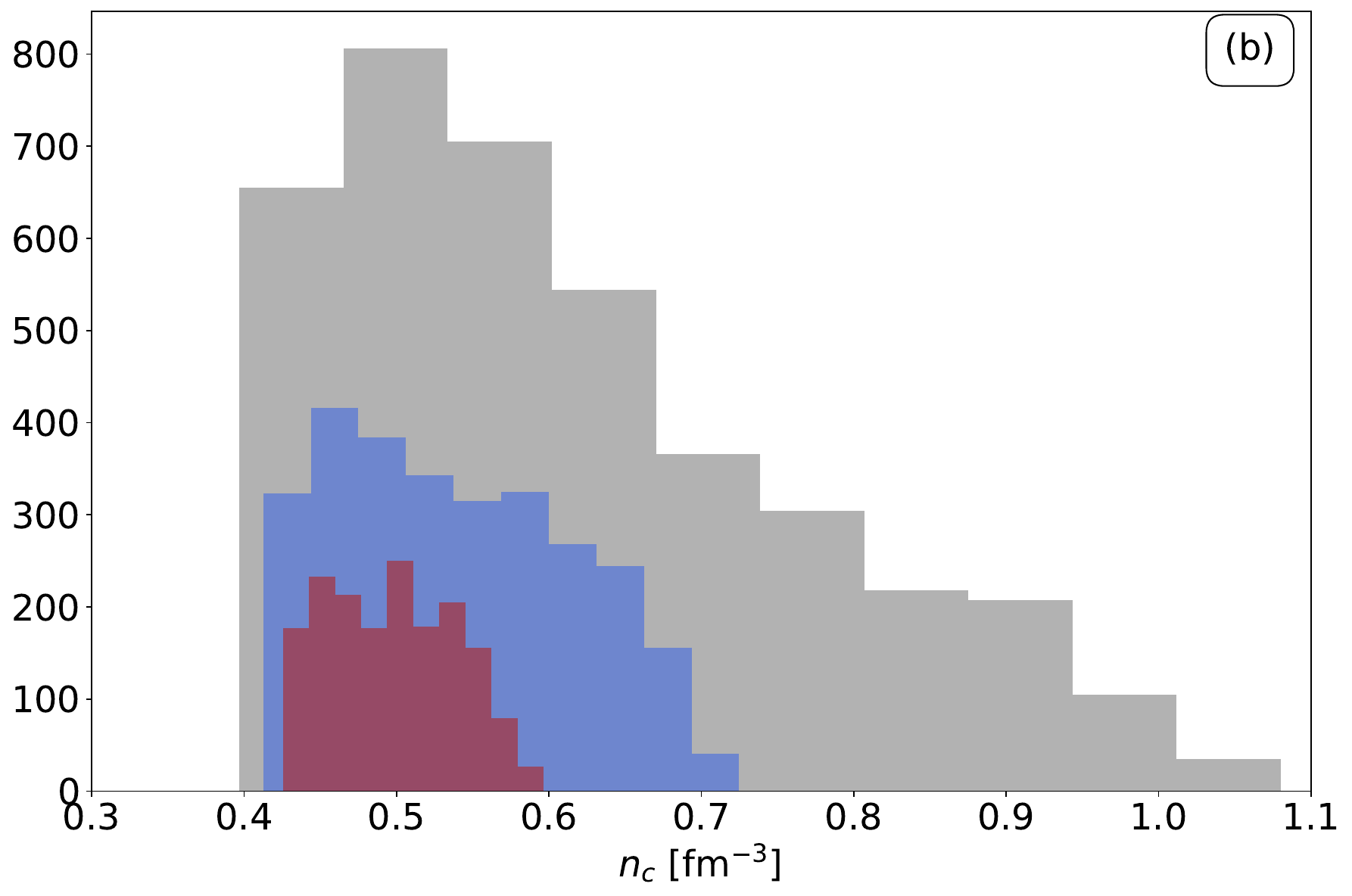}
\end{subfigure}
\caption{Relative central densities of $1.4 M_{\odot}$ and 2.0 $M_{\odot}$ in panels (a) and (b) respectively are reported for all selected models and adopting the same color code as in Figure \ref{fig:histo}.}    \label{fig:central_densities}
\end{figure}

The distributions of central densities are shown in Fig.~\ref{fig:central_densities} for an NS with $1.4M_\odot$ (a) and $2.0M_\odot$ (b). As expected the more massive NSs populate larger central densities. It is however interesting to remark that a large number of canonical mass NSs can have quite large central densities ($n_c\ge 3 n_\sat$) and some massive NSs can have rather low central densities ($n_c\approx 3 n_\sat$). For those cases in particular, 
it is possible that once introducing hyperons in our EOS, their threshold of appearance could occur at densities larger than the most massive NS's central densities. This could be a viable solution to the so-called hyperon puzzle, see e.g.\cite{Lonardoni:2014bwa}.

This occurs because the EOS can be soft or stiff: stiff EOS reach high masses for relatively low central densities (like the ones in red). Oppositely, soft EOSs (like most of the blue ones, compared to the red ones) have higher central densities for the same mass as stiff EOS.

Fig.~\ref{fig:central_densities} also shows that a better determination of the TOV mass can have an important impact on the prediction for central densities.

\subsection{Crust thickness}

Almost 300 sudden spin-up of the rotational frequency of pulsars have been observed since their discovery, for a review see Ref.~\cite{Espinoza:2011} and references therein. They are interpreted as abrupt transfer of angular momentum from the neutron superfluid to the solid crust, which is due to the unpinning of the superfluid vortices from the crystal lattice~\cite{Link:1999,Haskell:2015}. The large glitches observed in some pulsars, such as Vela for instance, require the crust to be thick enough to store a significant amount of angular momentum. In the present work, we study in more detail the crust thickness predicted by our selected models. 

Since our study is qualitative and we do not perform actual calculations for the crust, we adopt a simplified way to define the crust thickness, which is defined in the following way:
\begin{equation}
\Delta R_\mathrm{crust}=R-R(n=0.5n_\sat) \, ,
\end{equation}
where $R(n=0.5n_\sat)$ is the coordinate radius for the EOS for which the density is $n=0.5n_\sat$. It approximately locates the transition between the core and the crust.
The precise estimation of the crust thickness and crust moment of inertia is beyond the scope of the present study, but it would be interesting to study in more detail in future work. To do so, an accurate way to treat the crust is necessary by considering, for instance, a unified approach for the crust and the core, see Ref.~\cite{Fortin:2016,Carreau:2019} and references therein.

\begin{figure}[t]
\centering    
\begin{subfigure}[b]{0.46\textwidth}
\centering
\includegraphics[width=\textwidth]{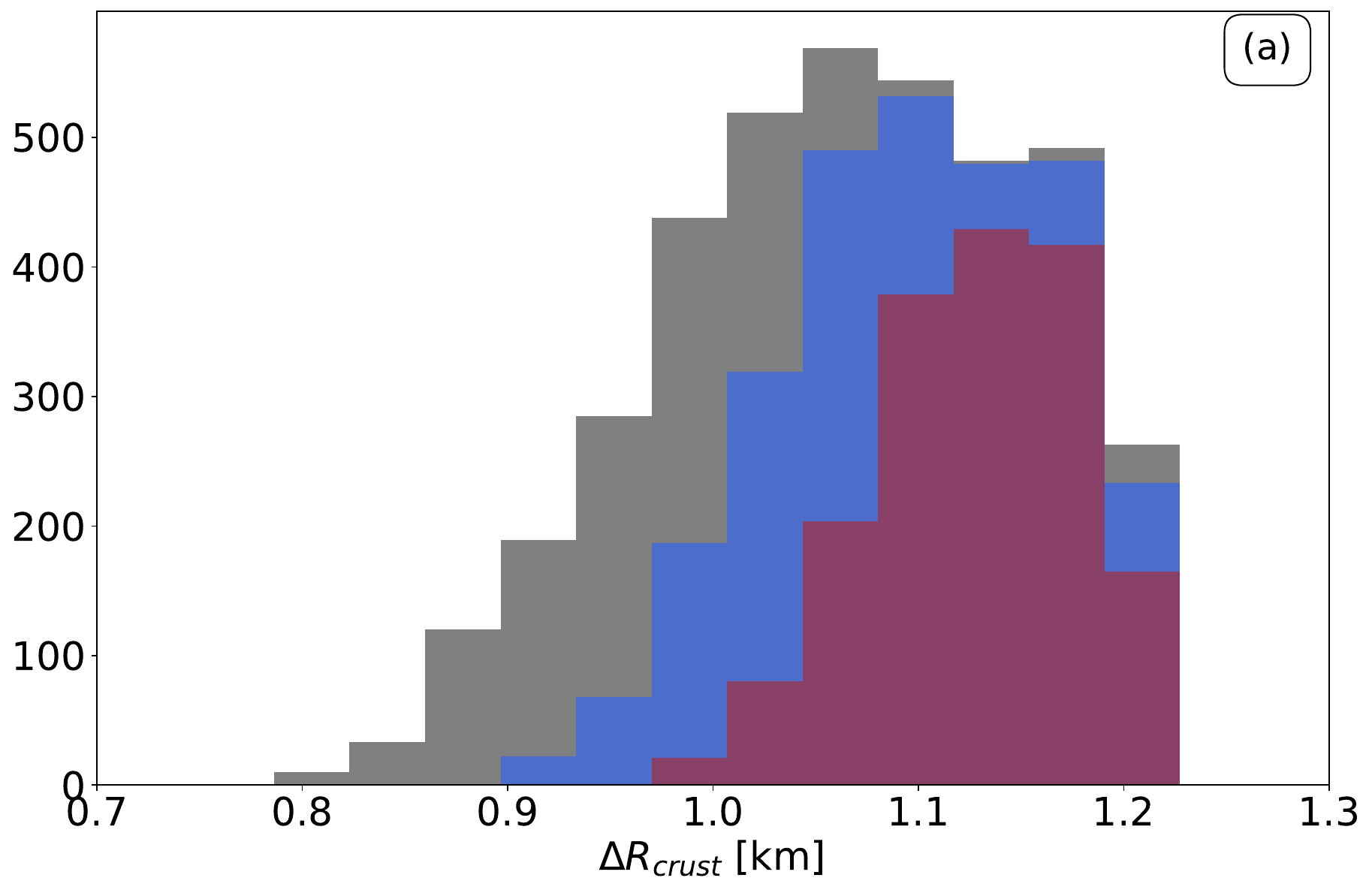} 
\end{subfigure}
\hfill
\begin{subfigure}[b]{0.45\textwidth}
\centering
\includegraphics[width=\textwidth]{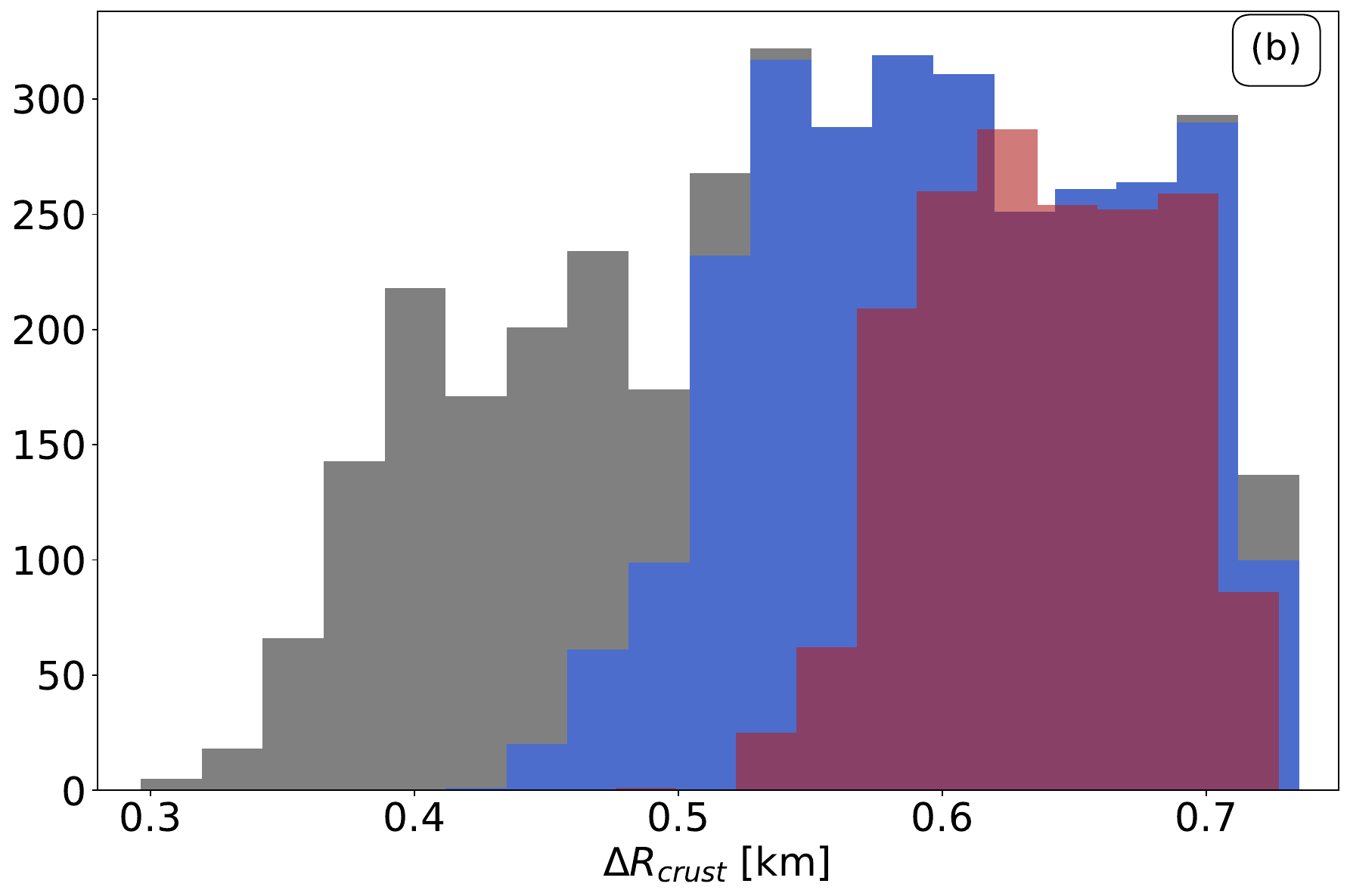}
\end{subfigure}
\caption{The crust thickness for $1.4 M_{\odot}$ and $2.0 M_{\odot}$ NS in panels (a) and (b) is reported respectively, for 68\% and 95\% confidence interval for the selected models. We have adopted the same color code as in Figure \ref{fig:histo}}
\label{fig:crust_thick}
\end{figure}

Fig.~\ref{fig:crust_thick} shows the distribution for the crust thickness for $1.4M_\odot$ (a) and $2.0 M_\odot$ (b) NSs obtained from the selected models. As expected the crust thickness decreases as the mass increases, because the gravitational force applied to the crust is larger for massive NS and squeezes the crust. In addition, it shows that a better determination of the TOV mass impacts directly the uncertainties in the crust thickness. Our 68\% selected models predict a crust thickness of $[1,1.2]$~km for $1.4M_\odot$ NS and $[0.52,0.72]$~km for $2.0M_\odot$ NS.

\subsection{Proton fraction}

Proton fraction is an important quantity that can trigger fast cooling of NS, through the direct URCA process. Indeed, it was shown in Ref.~\cite{Lattimer:1991} that the direct URCA process (dURCA) is quenched if the proton fraction satisfies $y_p<1/9$ (in the absence of muons). This condition is slightly modified in the presence of muons, but we will neglect the small contribution of muons in this discussion to keep it simple.

\begin{figure}[htbp]
\centering    
\begin{subfigure}[b]{0.45\textwidth}
\centering
\includegraphics[width=\textwidth]{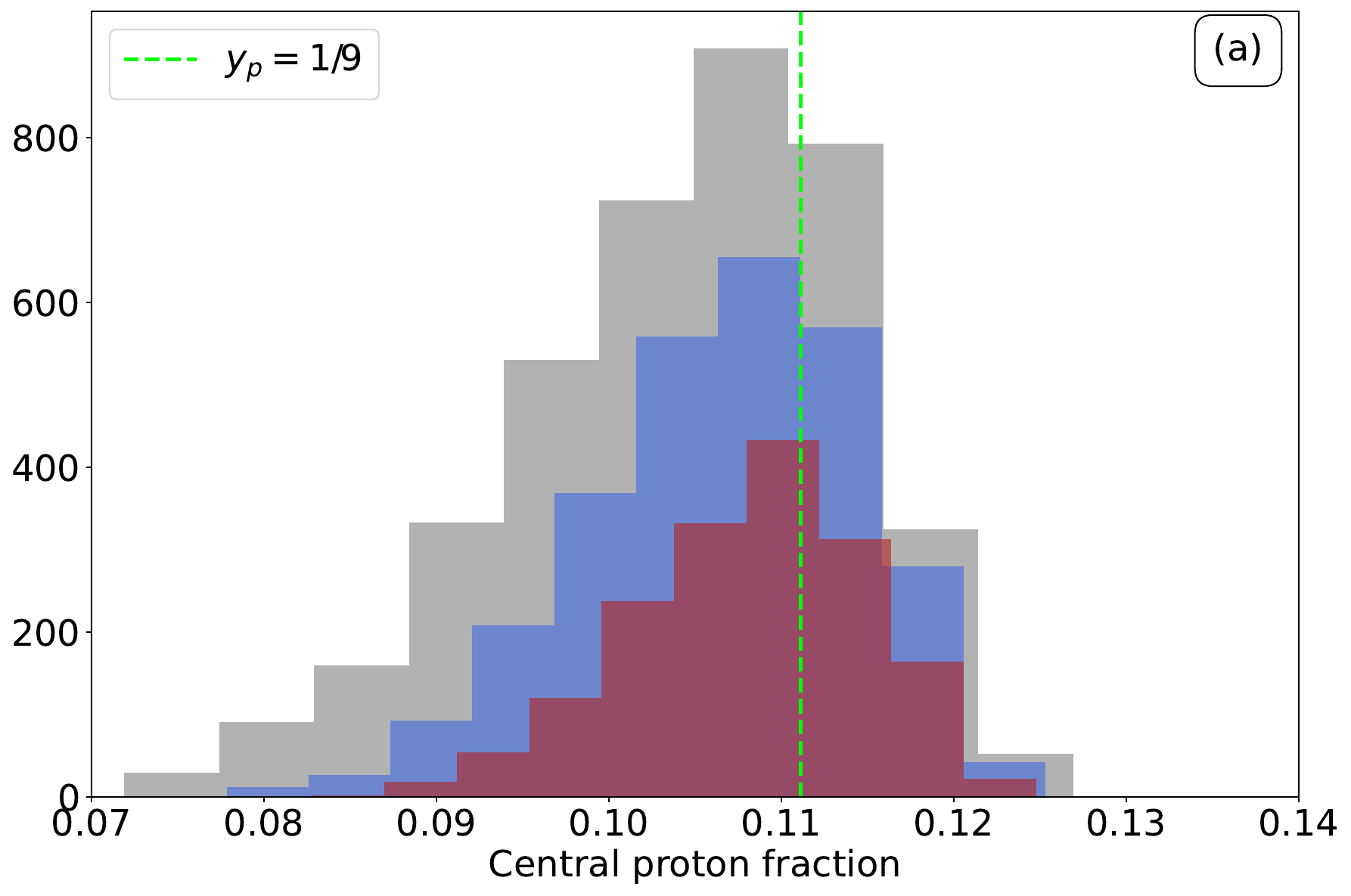}
\end{subfigure}
\hfill
\begin{subfigure}[b]{0.45\textwidth}
\centering
\includegraphics[width=\textwidth]{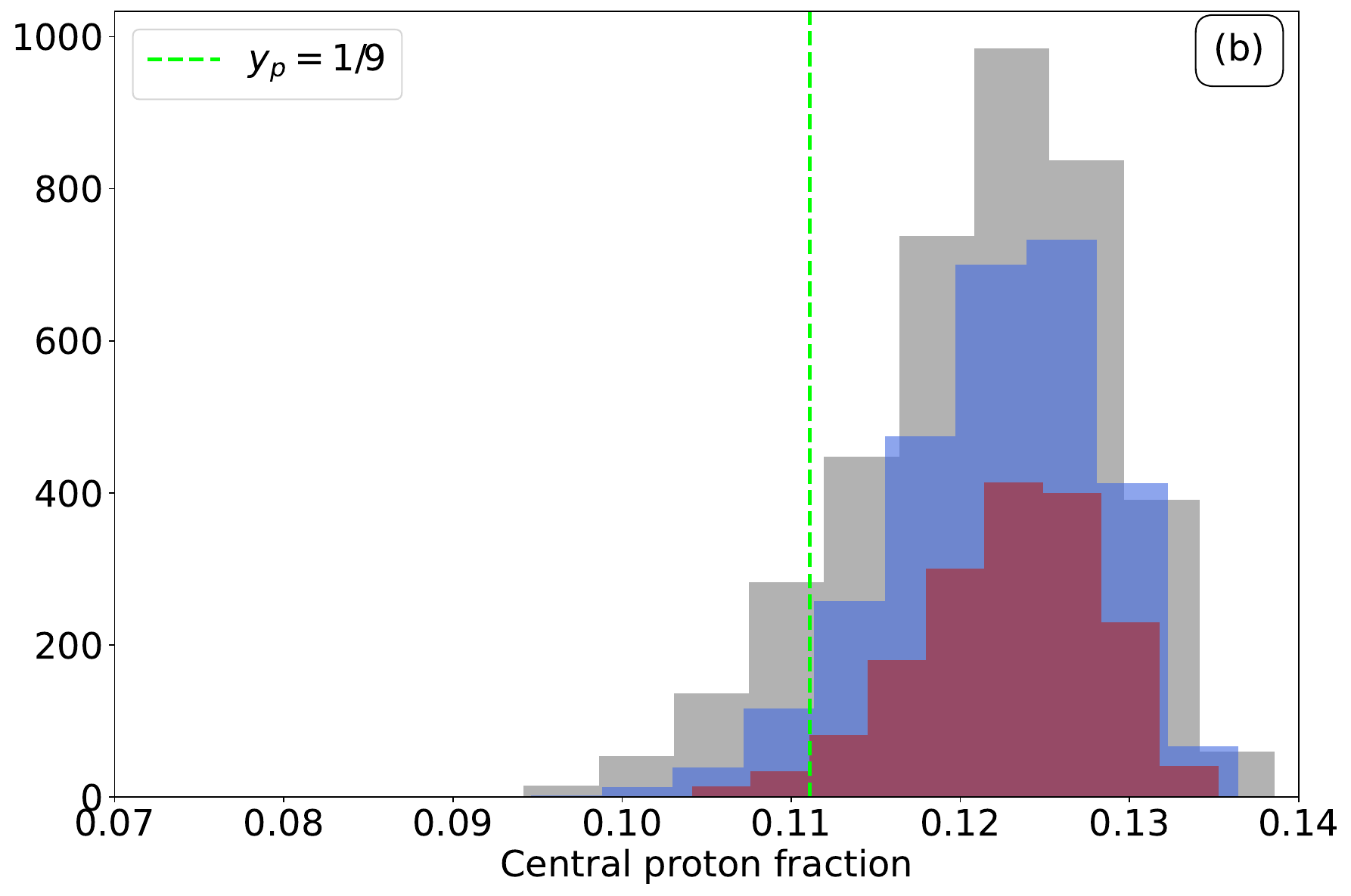}
\end{subfigure}
\caption{Central proton fraction for $1.4 M_{\odot}$ and $2.0 M_{\odot}$ NS in panels (a) and (b) respectively is reported, for 68\% and 95\% confidence interval for the selected models. The vertical dashed line (green) indicates the limit $y_p=1/9$. We have adopted the same color code as in Figure \ref{fig:histo}}
\label{fig:combined_plots_proton_fraction}
\end{figure}

Similarly to Fig.~\ref{fig:central_densities}, we represent in Fig.~\ref{fig:combined_plots_proton_fraction} the proton fraction at the center of NS for $1.4 M_{\odot}$ and $2.0 M_{\odot}$. The vertical dashed line represents the limit between the regions where dURCA is allowed (right) and where it is quenched (left). Interestingly, for canonical mass NS there are models for which the central proton fraction is above this limit, so where dURCA process is allowed, and models for which the central proton fraction is below this limit, so where dURCA is not allowed. Thus, the cooling of NS, which is mostly represented by a set of surface temperatures versus the estimated NS age, could be employed in order to further select the RMF models.
For massive NSs, most of the models predict that dURCA is allowed. This prediction weakly depends on the models, i.e., the different groups (colors) behave in a similar way.

In summary, we find that the number of selected models can be impacted by the inclusion of cooling data in the selection process. This can impact the predictions for canonical mass NSs. For massive NSs, all our models predict, however, a large proton fraction implying that dURCA fast cooling process is allowed.

\subsection{Symmetry energy}

The proton fraction is known to be determined by the density dependence of the symmetry energy. For this reason, we investigate in more detail the predictions for the symmetry energy at different densities based on the selected RMF models. In this section, we consider only three RMF models, PMP1, PMP2 and PMP3, since they represent well the small dispersion among the selected RMF models.

The symmetry energy is defined as
\begin{equation}
E_\sym(n_b) = E_\text{NM}/A(n_b)-E_\text{SM}/A(n_b)\, ,
\end{equation}
where $E_\text{NM}/A$ is the energy per particle in NM and $E_\text{SM}/A$ is the energy per particle in SM.
The energy per particle in AM can be expressed as an expansion in the isospin asymmetry parameter~\cite{Somasundaram:2021},
\begin{equation}
E/A(n_b,\delta) = E_\text{SM}/A(n_b) + E_{\sym,2}(n_b)\delta^2 + E_{\sym,4}(n_b)\delta^4 + \dots
\end{equation}
giving for the symmetry energy $E_\sym=E_{\sym,2}+E_{\sym,4}+\dots$ where $E_{\sym,2}$ is the quadratic contribution to the symmetry energy and $E_{\sym,4}$ the quartic contribution. More detail can be found in Ref.~\cite{Somasundaram:2021}.

Since it is expected that $E_{\sym,4}\ll E_{\sym,2}$, we have $E_\sym\approx E_{\sym,2}$: the symmetry energy and its quadratic contribution are often considered identical in the literature.

\begin{figure}[t]
\centering
\includegraphics[width=1\linewidth]{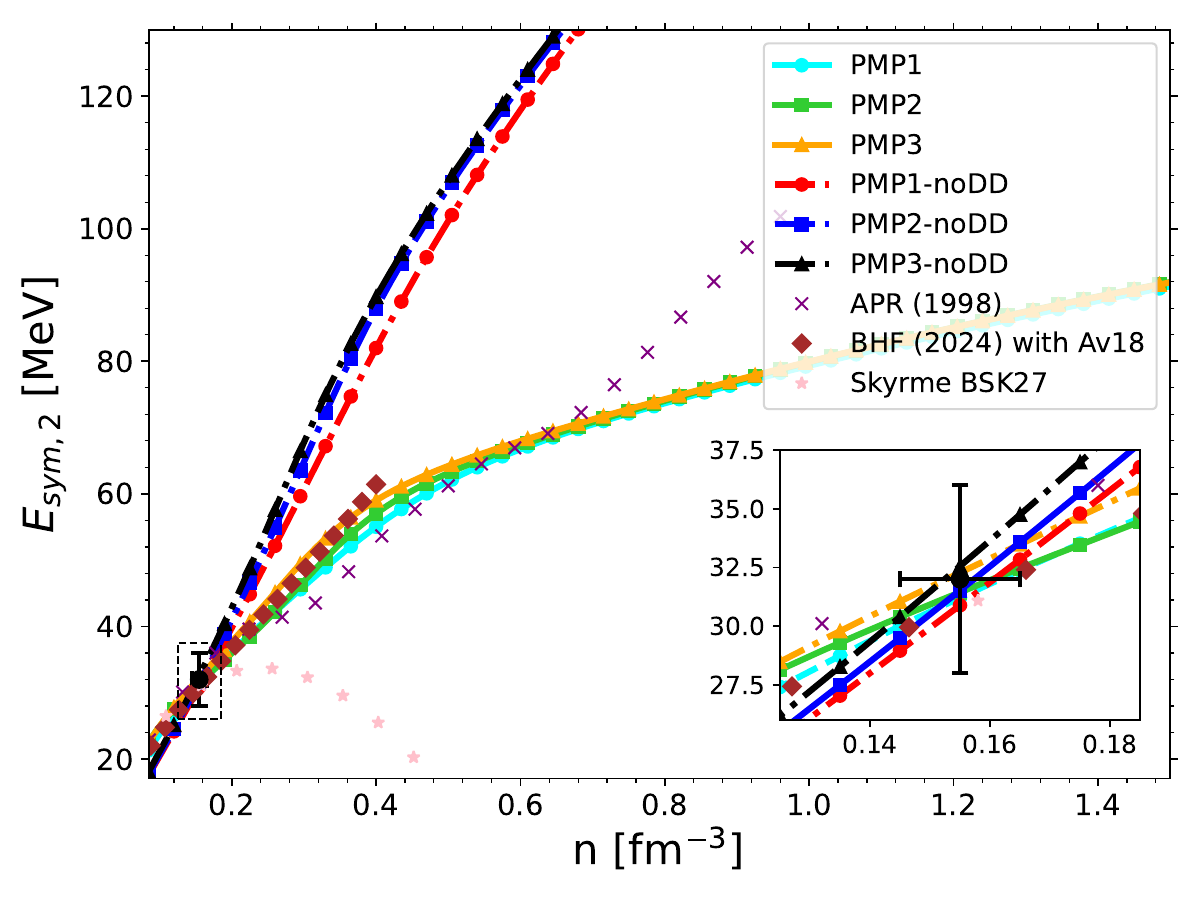}
\caption{The symmetry energy parameter $E_{\sym,2}$ as a function of the density for the models listed in Table \ref{tab:PMP1_NEP}, with and without the density dependence, labeled as noDD. The inset shows a detailed view of the symmetry energy around $n_\sat$ and the box represents the phenomenological uncertainties as given in Tab.~\ref{tab:nep_corrected_version}.}
\label{fig:Esym_plot}
\end{figure}

In Fig.~\ref{fig:Esym_plot}, the quadratic contribution to the symmetry energy $E_{\sym,2}$ predicted by PMP1, PMP2 and PMP3 is shown as a function of the nucleonic density $n_b$. As a function of $n_b$, the symmetry energy $E_{\sym,2}$ increases fast up to about $3n_\sat$ and then the slope is reduced (but it continues to increase). The change of the slope is due to the density-dependent coupling constant we have considered for the coupling between nucleon and $\rho$ meson, $g_{\rho}(n_b)$. To illustrate this, we plot in Fig.~\ref{fig:Esym_plot} the prediction for the symmetry energy $E_{\sym,2}$ based on the same PMP1, PMP2 and PMP3 models but where the nucleon-$\rho$ coupling constant is taken independent of the density and equal to its value at $n_\sat$ (dashed-dotted lines). Without density-dependent nucleon-$\rho$ coupling constant, the symmetry energy $E_{\sym,2}$ increases as a function of $n_b$ with a larger slope compared to the predictions for PMP1, PMP2 and PMP3, which include density-dependent nucleon-$\rho$ coupling constant.

We also compare the predictions of PMP1, PMP2 and PMP3 to microscopic approaches (APR-1998~\cite{APR:1998}, BHF with Av18 nuclear interaction~\cite{Vidana:2024}) and an example of phenomenological model (Skyrme BSk27~\cite{Goriely:2013}). The prediction for the symmetry energy $E_{\sym}$ for Skyrme BSk27 bends down above $n_\sat$, as many phenomenological models do, at variance with the microscopic predictions. The selected RMF models PMP1, PMP2 and PMP3 predict symmetry energies comparable to the microscopic approaches up to, at least, $\approx 3n_\sat$.

The inset in Fig.~\ref{fig:Esym_plot} shows that all models considered here (see the legend) are compatible with the symmetry energy $E_{\sym,2}$ at saturation density, for which the uncertainty is shown with a vertical line and the horizontal line represents the uncertainty in the saturation density.

\subsection{Core cooling}

The central proton fraction is shown in Fig.~\ref{fig:combined_plots_proton_fraction}, but a question, however, remains about the size of the central region where dURCA process is allowed or forbidden. 

\begin{figure}[t]
\centering
\begin{subfigure}[b]{0.45\textwidth}
\centering        \includegraphics[width=\textwidth]{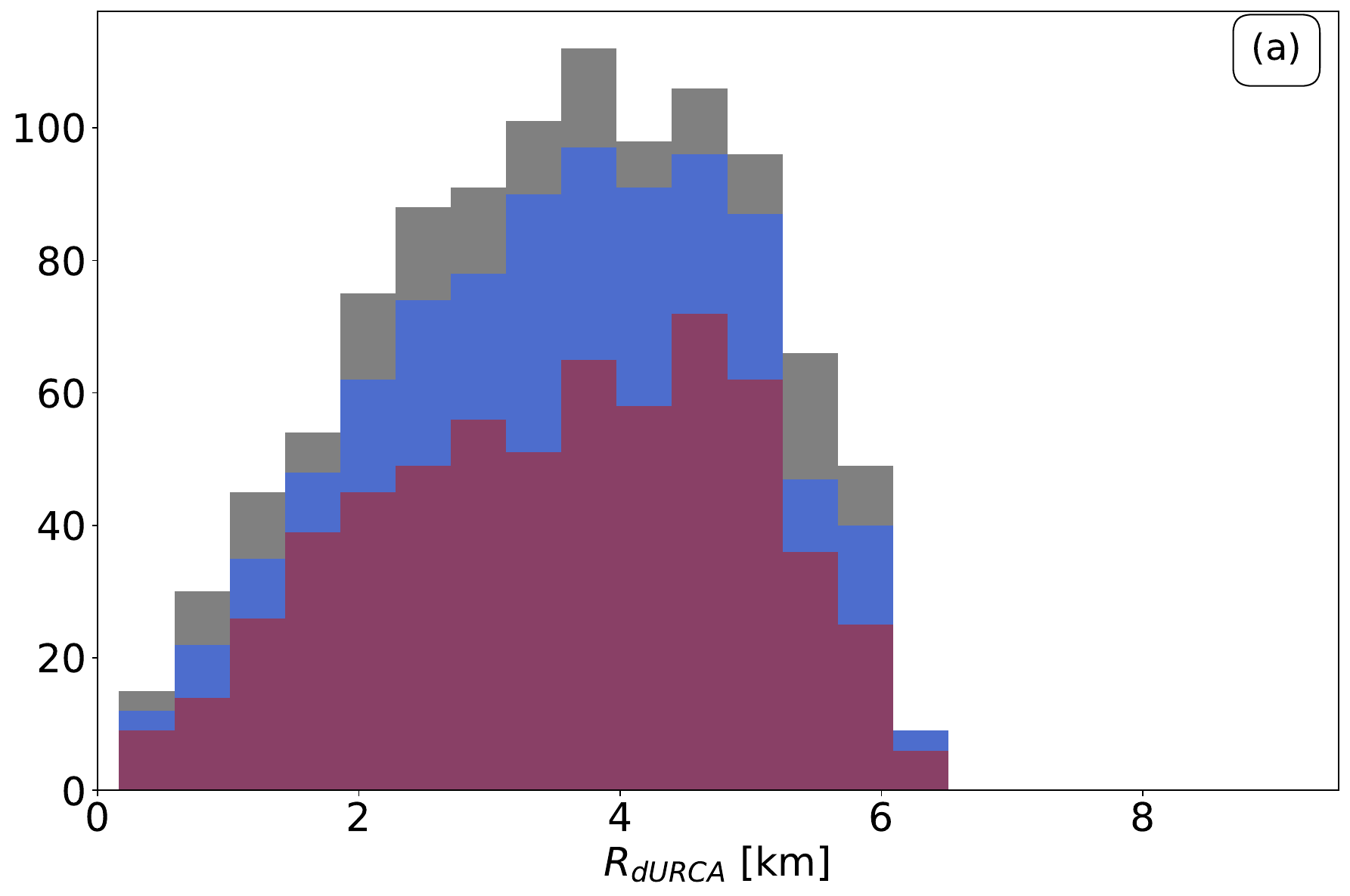} 
\end{subfigure}
\hfill
\begin{subfigure}[b]{0.45\textwidth}
\centering        \includegraphics[width=\textwidth]{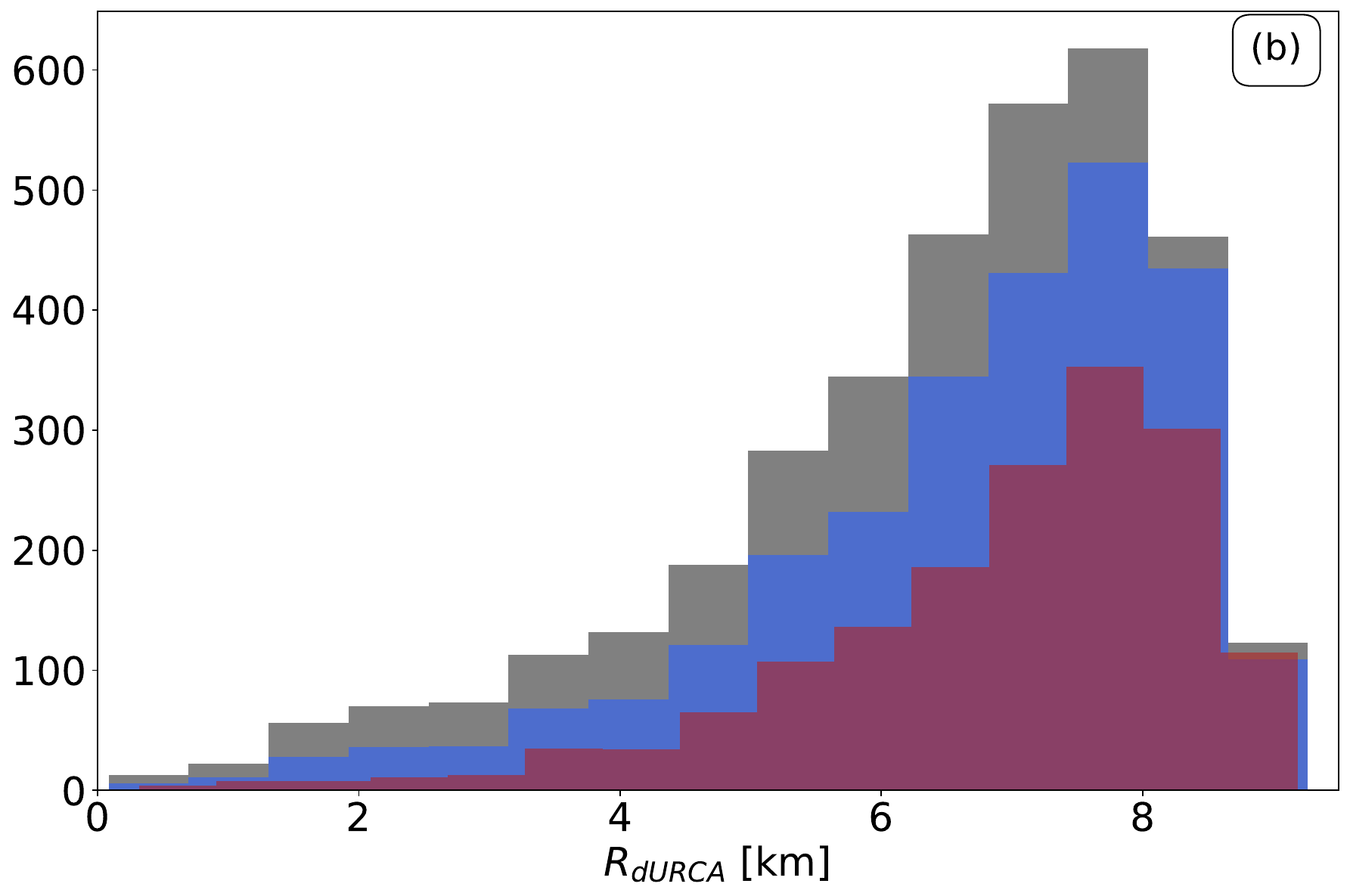} 
\end{subfigure}
\caption{The distribution of $R_{\mathrm{dURCA}}$ for $1.4 M_{\odot}$ and $2.0 M_{\odot}$ in panels (a) and (b) respectively are reported for all models shown in Figure \ref{fig:histo} for which $y_p>1/9$. See text for more details.}
\label{fig:core_cooling}
\end{figure}

In Fig.~\ref{fig:core_cooling}, we represent the distribution of models for the same two masses as before, $1.4 M_{\odot}$ and $2.0 M_{\odot}$, as a function of the fast cooling core radius $R_\mathrm{dURCA}$. The fast cooling core radius is defined as the coordinate radius inside which the proton fraction overpasses the limit $1/9$,
\begin{equation}
R_\mathrm{dURCA} = r_\tov(y_p=1/9) \, ,
\end{equation}
where $r_\tov$ is the solution of the TOV equations. 
For $1.4 M_{\odot}$ NS in panel (a) only the models with central proton fraction greater than $y_p=1/9$  are considered, the others lead to $R_\mathrm{dURCA}=0$~km.
Fig.~\ref{fig:core_cooling} informs us about the size of the core where the fast cooling dURCA process is allowed. The distribution of the quantity $R_\text{dURCA}$ is quite independent of the maximum mass (the different color regions). So even for the 68\% best models of the selection, it is possible to obtain EOS predicting a core size ranging inside $[0,6]$~km.

It is clear from Fig.~\ref{fig:core_cooling} that, for $1.4 M_{\odot}$ NS, 
it is possible to find a large core where the cooling is fast, impacting the entire cooling of the NS up to its surface. If these models are employed to predict cooling data and if they lead to results in contradiction with observations, all these models wil be excluded and we will be left with the other ones. Cooling data can therefore play an important role in the model selection.

\section{conclusions}

In this analysis, we have selected a large number of nucleonic RMF models constrained by nuclear physics and astrophysics: the selected models explore the range of NEPs given in Tab.~\ref{tab:nep_corrected_version} and the $\chi$EFT band shown in Fig.~\ref{fig:NM_binding_energy}, the tidal deformability extracted from GW170817 and the mass distribution shown in Fig.~\ref{fig:tov_mass}. By introducing a density dependence in the $\rho$ meson coupling constant, we obtain radii for canonical mass NS lower than the ones often predicted by usual RMF approaches.

The selected models are then employed to explore several properties of NS, using the large number of models as a way to explore the uncertainties in our predictions. They are compared to NICER masses-radii analyses for PSR J0030+0451 and PSR J0740+6620. We find that RMF models can be made soft enough to predict low values for neutron star radii compatible with GW170817 and, at larger densities, stiff enough to be compatible with NICER analyses for massive neutron star (PSR J0740+6620). The selected models are also compatible with NICER analyses for PSR J0030+0451. In addition, our models can also reach large values for the maximum mass, up to 2.6$M_\odot$. 

Overall, we find that the selection process plays an important role in determining neutron star properties. We have explored EOS properties and density dependence of the symmetry energy, the crust thickness, and predictions for core composition (proton fraction). The models suggest, for canonical mass NS, a large dispersion for the proton fraction around the dURCA threshold, suggesting that cooling data (observed surface temperature versus estimated age) could be efficiently employed in the selection process. For massive NSs, most of our models suggest a large proton fraction in the core allowing the dURCA fast cooling process. We have then discussed our results regarding the possibility for fast cooling of canonical mass NSs. We suggest several models, for which parameters are explicitly given, compatible with nuclear physics and astrophysical constraints, giving various sizes for the region where dURCA process is possible. 

In future investigations, we plan to include more degrees of freedom, e.g., hyperons, and to employ cooling data in the selection process.
\vskip 0.5cm

\textbf{Acknowledgments:}
We thank S. Vinciguerra for sharing with us the data points of the contours of PSR J0030+0451 of Ref.~\cite{Vinciguerra_2024}.
This study has been initiated during the internship of L.P. at IP2I Lyon supported by the CNRS-IN2P3 MAC masterproject, supporting J.M. as well. This work benefited from the support of the project RELANSE ANR-23-CE31-0027-01 of the French National Research Agency (ANR) and benefits from the LABEX Lyon Institute of Origins (ANR-10-LABX-0066). 



\appendix
\section{Parameters of the model}\label{Appendix_param}
One of the main advantage of the RMF model we use in this work
is the possibility to obtain analytical relations between 
the NEP and the free parameters appearing in the Lagrangian  \cite{Glendenning2000}. We report in the following those relations.

Let us start with the isoscalar vector channel. 
\begin{equation}
\Biggl( \frac{g_{\omega}}{m_{\omega}} \Biggr)^2 = \frac{m_N + E_{\sat} - \sqrt{k_{F_N}^2+m^{*2}_{D,\sat}}}{\nsat}
\end{equation}
Notice that in the field equations, only the ratio $g_\omega/m_\omega$ appears and therefore there is no need to specify which value for the mass of the $\omega$ meson, $m_\omega$, is adopted (similarly for the mass of the $\sigma$ meson, $m_\sigma$  and the mass of the $\rho$ meson, $m_\rho$).
Similarly for the isovector vector channel one obtains:

\begin{equation}
\Biggl( \frac{g_{\rho}(\nsat)}{m_{\rho}} \Biggr)^2 = \frac{ 8}{ \nsat} \Biggl[ E_{\sym,2} - \frac{k_{F_N}^2}{6 \sqrt{k_{F_N}^2 + m^{*2}_{D,\sat}}} \Biggr]
\end{equation}

The parameter $a_\rho$ allows to tune $L_\sym$ according to the following equation (obtained by deriving Eq. A2 with respect to $n_b$ and computing at $\nsat$):
\begin{eqnarray}
a_\rho&=&\frac{\nsat}{-2 (g_\rho(\nsat)/m_\rho)^2}(\frac{8}{\nsat^2}(L_\sym/3-E_{\sym,2}) \nonumber \\ &+&\frac{8}{\nsat^2} f-\frac{8}{\nsat}f')
\end{eqnarray}
where $f(k_{F_N},m^*_{D}(k_{F_N}))=\frac{k_{F_N}^2}{6 \sqrt{k_{F_N}^2 + m^{*2}_{D}}}$ and $f'(k_{F_N},m^*_{D}(k_{F_N}))=\frac{\mathrm{df}}{\mathrm{d}k_{F_N}}\frac{\mathrm{d}k_{F_N}}{\mathrm{d}n_b}$ its derivative with respect to the baryon density (computed at saturation). For computing $f'(k_{F_N},m^*_{D}(k_{F_N}))$ one needs the derivative of the effective mass $m^*_{D} $ with respect to $k_{F_N}$ which in turn is proportional to the derivative of the $\sigma$ field that can be computed analitically from the mean field equation (\ref{eq:sigma}).

The three parameters for the isoscalar scalar channel are obtained by the following equations:

\begin{subequations}
\label{indice 1 4.191 Glen.}
\begin{align}
\alpha_1&= K_\sat - \Biggl(\frac{g_{\omega}}{m_{\omega}} \Biggr)^2 \frac{6k_{F_N}^3}{\pi ^2} - \frac{3 k_{F_N}^2}{\sqrt{k_{F_N}^2 + m^{*2}_{D,\sat}}}, \\
\beta_1 & =2\alpha_1 (m_N-m^{*}_{D,\sat})m_N, \\
\gamma_1&=3 \alpha_1 (m_N-m^{*}_{D,\sat})^2, \\
\delta_1 &= - \alpha_1 I_1- \frac{6 k_{F_N}^3}{\pi^2} \Biggl( \frac{m^{*}_{D,\sat}}{\sqrt{k_{F_N}^2+m^{*2}_{D,\sat}}}\Biggr)^2
\end{align}
\end{subequations}

\begin{subequations}
\label{indice 2 4.192 Glen.}
\begin{align}
\alpha_2&= \frac{1}{2} (m_N-m^{*}_{D,\sat})^2, \\
\beta_2 & =\frac{1}{3} m_N (m_N-m^{*}_{D,\sat})^3, \\
\gamma_2&=\frac{1}{4} (m_N-m^{*}_{D,\sat})^4, \\
\delta_2 &= \nsat (m_N + E_\sat) - I_2 - \frac{1}{2} \bigg(\frac{g_{\omega}}{m_{\omega}}\bigg)^2 \nsat ^2
\end{align}
\end{subequations}

\begin{subequations}
\label{indice 3 4.193 Glen.}
\begin{align}
\alpha_3&= m_N-m^{*}_{D,\sat}, \\
\beta_3 & = m_N (m_N-m^{*}_{D,\sat})^2, \\
\gamma_3&= (m_N-m^{*}_{D,\sat})^3, \\
\delta_3 &= I_3
\end{align}
\end{subequations}
where
\begin{subequations}
\begin{align}
x &= k_{F_N}/m^{*}_{D,\sat}, \\
t &= \sqrt{1+x^2},\\
I_1&= \frac{2}{\pi^2} m^{*2}_{D,sat} \Biggl[\frac{1}{2} x t + \frac{x}{t} -\frac{3}{2}\ln(x+t) \Biggr],\\
I_2&= \frac{2}{\pi^2} m^{*4}_{D,\sat} \frac{1}{4}\Biggl[ x t^3 - \frac{1}{2} xt -\frac{1}{2} \ln(x+t)\Biggr],\\
I_3&= \frac{2}{\pi^2} m^{*3}_{D,\sat} \frac{1}{2}\Biggl[ x t - \ln(x+t)\Biggr]    
\end{align}
\end{subequations}

Finally:

\begin{equation}
c =  \frac{\alpha_3 \beta_2 \delta_1-\alpha_2 \beta_3 \delta_1-\alpha_3 \beta_1 \delta_2+\alpha_1 \beta_3 \delta_2+\alpha_2 \beta_1 \delta_3-\alpha_1 \beta_2 \delta_3}{\alpha_3 \beta_2 \gamma_1-\alpha_2 \beta_3 \gamma_1-\alpha_3 \beta_1 \gamma_2+\alpha_1 \beta_3 \gamma_2+\alpha_2 \beta_1 \gamma_3-\alpha_1 \beta_2 \gamma_3}
\end{equation}

\begin{equation}
b = \frac{(\alpha_2 \delta_1 - \alpha_1 \delta_2) - (\alpha_2 \gamma_1-\alpha_1 \gamma_2)c}{\alpha_2 \beta_1-\alpha_1 \beta_2}  
\end{equation}

\begin{equation}
\Biggl( \frac{g_{\sigma}}{m_{\sigma}} \Biggr)^2 = \frac{ \alpha_1}{\delta_1 -\gamma_1 c-\beta_1 b}
\end{equation}

\begin{thebibliography}{56}%
\makeatletter
\providecommand \@ifxundefined [1]{%
 \@ifx{#1\undefined}
}%
\providecommand \@ifnum [1]{%
 \ifnum #1\expandafter \@firstoftwo
 \else \expandafter \@secondoftwo
 \fi
}%
\providecommand \@ifx [1]{%
 \ifx #1\expandafter \@firstoftwo
 \else \expandafter \@secondoftwo
 \fi
}%
\providecommand \natexlab [1]{#1}%
\providecommand \enquote  [1]{``#1''}%
\providecommand \bibnamefont  [1]{#1}%
\providecommand \bibfnamefont [1]{#1}%
\providecommand \citenamefont [1]{#1}%
\providecommand \href@noop [0]{\@secondoftwo}%
\providecommand \href [0]{\begingroup \@sanitize@url \@href}%
\providecommand \@href[1]{\@@startlink{#1}\@@href}%
\providecommand \@@href[1]{\endgroup#1\@@endlink}%
\providecommand \@sanitize@url [0]{\catcode `\\12\catcode `\$12\catcode `\&12\catcode `\#12\catcode `\^12\catcode `\_12\catcode `\%12\relax}%
\providecommand \@@startlink[1]{}%
\providecommand \@@endlink[0]{}%
\providecommand \url  [0]{\begingroup\@sanitize@url \@url }%
\providecommand \@url [1]{\endgroup\@href {#1}{\urlprefix }}%
\providecommand \urlprefix  [0]{URL }%
\providecommand \Eprint [0]{\href }%
\providecommand \doibase [0]{http://dx.doi.org/}%
\providecommand \selectlanguage [0]{\@gobble}%
\providecommand \bibinfo  [0]{\@secondoftwo}%
\providecommand \bibfield  [0]{\@secondoftwo}%
\providecommand \translation [1]{[#1]}%
\providecommand \BibitemOpen [0]{}%
\providecommand \bibitemStop [0]{}%
\providecommand \bibitemNoStop [0]{.\EOS\space}%
\providecommand \EOS [0]{\spacefactor3000\relax}%
\providecommand \BibitemShut  [1]{\csname bibitem#1\endcsname}%
\let\auto@bib@innerbib\@empty
\bibitem [{\citenamefont {Baldo}\ \emph {et~al.}(1997)\citenamefont {Baldo}, \citenamefont {Bombaci},\ and\ \citenamefont {Burgio}}]{Baldo:1997ag}%
  \BibitemOpen
  \bibfield  {author} {\bibinfo {author} {\bibfnamefont {M.}~\bibnamefont {Baldo}}, \bibinfo {author} {\bibfnamefont {I.}~\bibnamefont {Bombaci}}, \ and\ \bibinfo {author} {\bibfnamefont {G.~F.}\ \bibnamefont {Burgio}},\ }\href@noop {} {\bibfield  {journal} {\bibinfo  {journal} {Astron. Astrophys.}\ }\textbf {\bibinfo {volume} {328}},\ \bibinfo {pages} {274} (\bibinfo {year} {1997})},\ \Eprint {http://arxiv.org/abs/astro-ph/9707277} {arXiv:astro-ph/9707277} \BibitemShut {NoStop}%
\bibitem [{\citenamefont {Bogner}\ \emph {et~al.}(2010)\citenamefont {Bogner}, \citenamefont {Furnstahl},\ and\ \citenamefont {Schwenk}}]{Bogner:2009bt}%
  \BibitemOpen
  \bibfield  {author} {\bibinfo {author} {\bibfnamefont {S.~K.}\ \bibnamefont {Bogner}}, \bibinfo {author} {\bibfnamefont {R.~J.}\ \bibnamefont {Furnstahl}}, \ and\ \bibinfo {author} {\bibfnamefont {A.}~\bibnamefont {Schwenk}},\ }\href {\doibase 10.1016/j.ppnp.2010.03.001} {\bibfield  {journal} {\bibinfo  {journal} {Prog. Part. Nucl. Phys.}\ }\textbf {\bibinfo {volume} {65}},\ \bibinfo {pages} {94} (\bibinfo {year} {2010})},\ \Eprint {http://arxiv.org/abs/0912.3688} {arXiv:0912.3688 [nucl-th]} \BibitemShut {NoStop}%
\bibitem [{\citenamefont {Carlson}\ \emph {et~al.}(2015)\citenamefont {Carlson}, \citenamefont {Gandolfi}, \citenamefont {Pederiva}, \citenamefont {Pieper}, \citenamefont {Schiavilla}, \citenamefont {Schmidt},\ and\ \citenamefont {Wiringa}}]{Carlson:2014vla}%
  \BibitemOpen
  \bibfield  {author} {\bibinfo {author} {\bibfnamefont {J.}~\bibnamefont {Carlson}}, \bibinfo {author} {\bibfnamefont {S.}~\bibnamefont {Gandolfi}}, \bibinfo {author} {\bibfnamefont {F.}~\bibnamefont {Pederiva}}, \bibinfo {author} {\bibfnamefont {S.~C.}\ \bibnamefont {Pieper}}, \bibinfo {author} {\bibfnamefont {R.}~\bibnamefont {Schiavilla}}, \bibinfo {author} {\bibfnamefont {K.~E.}\ \bibnamefont {Schmidt}}, \ and\ \bibinfo {author} {\bibfnamefont {R.~B.}\ \bibnamefont {Wiringa}},\ }\href {\doibase 10.1103/RevModPhys.87.1067} {\bibfield  {journal} {\bibinfo  {journal} {Rev. Mod. Phys.}\ }\textbf {\bibinfo {volume} {87}},\ \bibinfo {pages} {1067} (\bibinfo {year} {2015})},\ \Eprint {http://arxiv.org/abs/1412.3081} {arXiv:1412.3081 [nucl-th]} \BibitemShut {NoStop}%
\bibitem [{\citenamefont {Mueller}\ and\ \citenamefont {Serot}(1996)}]{Mueller:1996pm}%
  \BibitemOpen
  \bibfield  {author} {\bibinfo {author} {\bibfnamefont {H.}~\bibnamefont {Mueller}}\ and\ \bibinfo {author} {\bibfnamefont {B.~D.}\ \bibnamefont {Serot}},\ }\href {\doibase 10.1016/0375-9474(96)00187-X} {\bibfield  {journal} {\bibinfo  {journal} {Nucl. Phys. A}\ }\textbf {\bibinfo {volume} {606}},\ \bibinfo {pages} {508} (\bibinfo {year} {1996})},\ \Eprint {http://arxiv.org/abs/nucl-th/9603037} {arXiv:nucl-th/9603037} \BibitemShut {NoStop}%
\bibitem [{\citenamefont {Margueron}\ \emph {et~al.}(2009)\citenamefont {Margueron}, \citenamefont {Goriely}, \citenamefont {Grasso}, \citenamefont {Colo},\ and\ \citenamefont {Sagawa}}]{Margueron:2009jf}%
  \BibitemOpen
  \bibfield  {author} {\bibinfo {author} {\bibfnamefont {J.}~\bibnamefont {Margueron}}, \bibinfo {author} {\bibfnamefont {S.}~\bibnamefont {Goriely}}, \bibinfo {author} {\bibfnamefont {M.}~\bibnamefont {Grasso}}, \bibinfo {author} {\bibfnamefont {G.}~\bibnamefont {Colo}}, \ and\ \bibinfo {author} {\bibfnamefont {H.}~\bibnamefont {Sagawa}},\ }\href {\doibase 10.1088/0954-3899/36/12/125103} {\bibfield  {journal} {\bibinfo  {journal} {J. Phys. G}\ }\textbf {\bibinfo {volume} {36}},\ \bibinfo {pages} {125103} (\bibinfo {year} {2009})},\ \Eprint {http://arxiv.org/abs/0906.3508} {arXiv:0906.3508 [nucl-th]} \BibitemShut {NoStop}%
\bibitem [{\citenamefont {Steiner}\ \emph {et~al.}(2010)\citenamefont {Steiner}, \citenamefont {Lattimer},\ and\ \citenamefont {Brown}}]{Steiner:2010fz}%
  \BibitemOpen
  \bibfield  {author} {\bibinfo {author} {\bibfnamefont {A.~W.}\ \bibnamefont {Steiner}}, \bibinfo {author} {\bibfnamefont {J.~M.}\ \bibnamefont {Lattimer}}, \ and\ \bibinfo {author} {\bibfnamefont {E.~F.}\ \bibnamefont {Brown}},\ }\href {\doibase 10.1088/0004-637X/722/1/33} {\bibfield  {journal} {\bibinfo  {journal} {Astrophys. J.}\ }\textbf {\bibinfo {volume} {722}},\ \bibinfo {pages} {33} (\bibinfo {year} {2010})},\ \Eprint {http://arxiv.org/abs/1005.0811} {arXiv:1005.0811 [astro-ph.HE]} \BibitemShut {NoStop}%
\bibitem [{\citenamefont {Margueron}\ \emph {et~al.}(2018{\natexlab{a}})\citenamefont {Margueron}, \citenamefont {Hoffmann~Casali},\ and\ \citenamefont {Gulminelli}}]{Margueron:2017lup}%
  \BibitemOpen
  \bibfield  {author} {\bibinfo {author} {\bibfnamefont {J.}~\bibnamefont {Margueron}}, \bibinfo {author} {\bibfnamefont {R.}~\bibnamefont {Hoffmann~Casali}}, \ and\ \bibinfo {author} {\bibfnamefont {F.}~\bibnamefont {Gulminelli}},\ }\href {\doibase 10.1103/PhysRevC.97.025806} {\bibfield  {journal} {\bibinfo  {journal} {Phys. Rev. C}\ }\textbf {\bibinfo {volume} {97}},\ \bibinfo {pages} {025806} (\bibinfo {year} {2018}{\natexlab{a}})},\ \Eprint {http://arxiv.org/abs/1708.06895} {arXiv:1708.06895 [nucl-th]} \BibitemShut {NoStop}%
\bibitem [{\citenamefont {Margueron}\ \emph {et~al.}(2018{\natexlab{b}})\citenamefont {Margueron}, \citenamefont {Hoffmann~Casali},\ and\ \citenamefont {Gulminelli}}]{Margueron:2017eqc}%
  \BibitemOpen
  \bibfield  {author} {\bibinfo {author} {\bibfnamefont {J.}~\bibnamefont {Margueron}}, \bibinfo {author} {\bibfnamefont {R.}~\bibnamefont {Hoffmann~Casali}}, \ and\ \bibinfo {author} {\bibfnamefont {F.}~\bibnamefont {Gulminelli}},\ }\href {\doibase 10.1103/PhysRevC.97.025805} {\bibfield  {journal} {\bibinfo  {journal} {Phys. Rev. C}\ }\textbf {\bibinfo {volume} {97}},\ \bibinfo {pages} {025805} (\bibinfo {year} {2018}{\natexlab{b}})},\ \Eprint {http://arxiv.org/abs/1708.06894} {arXiv:1708.06894 [nucl-th]} \BibitemShut {NoStop}%
\bibitem [{\citenamefont {Drischler}\ \emph {et~al.}(2020{\natexlab{a}})\citenamefont {Drischler}, \citenamefont {Furnstahl}, \citenamefont {Melendez},\ and\ \citenamefont {Phillips}}]{Drischler:2020hwi}%
  \BibitemOpen
  \bibfield  {author} {\bibinfo {author} {\bibfnamefont {C.}~\bibnamefont {Drischler}}, \bibinfo {author} {\bibfnamefont {R.~J.}\ \bibnamefont {Furnstahl}}, \bibinfo {author} {\bibfnamefont {J.~A.}\ \bibnamefont {Melendez}}, \ and\ \bibinfo {author} {\bibfnamefont {D.~R.}\ \bibnamefont {Phillips}},\ }\href {\doibase 10.1103/PhysRevLett.125.202702} {\bibfield  {journal} {\bibinfo  {journal} {Phys. Rev. Lett.}\ }\textbf {\bibinfo {volume} {125}},\ \bibinfo {pages} {202702} (\bibinfo {year} {2020}{\natexlab{a}})},\ \Eprint {http://arxiv.org/abs/2004.07232} {arXiv:2004.07232 [nucl-th]} \BibitemShut {NoStop}%
\bibitem [{\citenamefont {Traversi}\ \emph {et~al.}(2020)\citenamefont {Traversi}, \citenamefont {Char},\ and\ \citenamefont {Pagliara}}]{Traversi:2020aaa}%
  \BibitemOpen
  \bibfield  {author} {\bibinfo {author} {\bibfnamefont {S.}~\bibnamefont {Traversi}}, \bibinfo {author} {\bibfnamefont {P.}~\bibnamefont {Char}}, \ and\ \bibinfo {author} {\bibfnamefont {G.}~\bibnamefont {Pagliara}},\ }\href {\doibase 10.3847/1538-4357/ab99c1} {\bibfield  {journal} {\bibinfo  {journal} {Astrophys. J.}\ }\textbf {\bibinfo {volume} {897}},\ \bibinfo {pages} {165} (\bibinfo {year} {2020})},\ \Eprint {http://arxiv.org/abs/2002.08951} {arXiv:2002.08951 [astro-ph.HE]} \BibitemShut {NoStop}%
\bibitem [{\citenamefont {Alvarez-Castillo}\ \emph {et~al.}(2020)\citenamefont {Alvarez-Castillo}, \citenamefont {Ayriyan}, \citenamefont {Barnaföldi}, \citenamefont {Grigorian},\ and\ \citenamefont {Pósfay}}]{Alvarez-Castillo:2020}%
  \BibitemOpen
  \bibfield  {author} {\bibinfo {author} {\bibfnamefont {D.}~\bibnamefont {Alvarez-Castillo}}, \bibinfo {author} {\bibfnamefont {A.}~\bibnamefont {Ayriyan}}, \bibinfo {author} {\bibfnamefont {G.~G.}\ \bibnamefont {Barnaföldi}}, \bibinfo {author} {\bibfnamefont {H.}~\bibnamefont {Grigorian}}, \ and\ \bibinfo {author} {\bibfnamefont {P.}~\bibnamefont {Pósfay}},\ }\href {\doibase 10.1140/epjst/e2020-000106-4} {\bibfield  {journal} {\bibinfo  {journal} {Eur. Phys. J. Spec. Top.}\ }\textbf {\bibinfo {volume} {229}},\ \bibinfo {pages} {3615} (\bibinfo {year} {2020})}\BibitemShut {NoStop}%
\bibitem [{\citenamefont {Malik}\ \emph {et~al.}(2022)\citenamefont {Malik}, \citenamefont {Ferreira}, \citenamefont {Agrawal},\ and\ \citenamefont {Provid\^encia}}]{Malik:2022zol}%
  \BibitemOpen
  \bibfield  {author} {\bibinfo {author} {\bibfnamefont {T.}~\bibnamefont {Malik}}, \bibinfo {author} {\bibfnamefont {M.}~\bibnamefont {Ferreira}}, \bibinfo {author} {\bibfnamefont {B.~K.}\ \bibnamefont {Agrawal}}, \ and\ \bibinfo {author} {\bibfnamefont {C.}~\bibnamefont {Provid\^encia}},\ }\href {\doibase 10.3847/1538-4357/ac5d3c} {\bibfield  {journal} {\bibinfo  {journal} {Astrophys. J.}\ }\textbf {\bibinfo {volume} {930}},\ \bibinfo {pages} {17} (\bibinfo {year} {2022})},\ \Eprint {http://arxiv.org/abs/2201.12552} {arXiv:2201.12552 [nucl-th]} \BibitemShut {NoStop}%
\bibitem [{\citenamefont {Zhu}\ \emph {et~al.}(2023)\citenamefont {Zhu}, \citenamefont {Li},\ and\ \citenamefont {Liu}}]{Zhu:2023}%
  \BibitemOpen
  \bibfield  {author} {\bibinfo {author} {\bibfnamefont {Z.}~\bibnamefont {Zhu}}, \bibinfo {author} {\bibfnamefont {A.}~\bibnamefont {Li}}, \ and\ \bibinfo {author} {\bibfnamefont {T.}~\bibnamefont {Liu}},\ }\href {\doibase 10.3847/1538-4357/acac1f} {\bibfield  {journal} {\bibinfo  {journal} {The Astrophysical Journal}\ }\textbf {\bibinfo {volume} {943}},\ \bibinfo {pages} {163} (\bibinfo {year} {2023})}\BibitemShut {NoStop}%
\bibitem [{\citenamefont {Beznogov}\ and\ \citenamefont {Raduta}(2023)}]{Beznogov:2023}%
  \BibitemOpen
  \bibfield  {author} {\bibinfo {author} {\bibfnamefont {M.~V.}\ \bibnamefont {Beznogov}}\ and\ \bibinfo {author} {\bibfnamefont {A.~R.}\ \bibnamefont {Raduta}},\ }\href {\doibase 10.1103/PhysRevC.107.045803} {\bibfield  {journal} {\bibinfo  {journal} {Phys. Rev. C}\ }\textbf {\bibinfo {volume} {107}},\ \bibinfo {pages} {045803} (\bibinfo {year} {2023})}\BibitemShut {NoStop}%
\bibitem [{\citenamefont {Guo}\ \emph {et~al.}(2024)\citenamefont {Guo}, \citenamefont {Xiong}, \citenamefont {Ma},\ and\ \citenamefont {Ma}}]{Guo:2023mhf}%
  \BibitemOpen
  \bibfield  {author} {\bibinfo {author} {\bibfnamefont {L.-J.}\ \bibnamefont {Guo}}, \bibinfo {author} {\bibfnamefont {J.-Y.}\ \bibnamefont {Xiong}}, \bibinfo {author} {\bibfnamefont {Y.}~\bibnamefont {Ma}}, \ and\ \bibinfo {author} {\bibfnamefont {Y.-L.}\ \bibnamefont {Ma}},\ }\href {\doibase 10.3847/1538-4357/ad2e8d} {\bibfield  {journal} {\bibinfo  {journal} {Astrophys. J.}\ }\textbf {\bibinfo {volume} {965}},\ \bibinfo {pages} {47} (\bibinfo {year} {2024})},\ \Eprint {http://arxiv.org/abs/2309.11227} {arXiv:2309.11227 [nucl-th]} \BibitemShut {NoStop}%
\bibitem [{\citenamefont {Drago}\ \emph {et~al.}(2014)\citenamefont {Drago}, \citenamefont {Lavagno}, \citenamefont {Pagliara},\ and\ \citenamefont {Pigato}}]{Drago:2014oja}%
  \BibitemOpen
  \bibfield  {author} {\bibinfo {author} {\bibfnamefont {A.}~\bibnamefont {Drago}}, \bibinfo {author} {\bibfnamefont {A.}~\bibnamefont {Lavagno}}, \bibinfo {author} {\bibfnamefont {G.}~\bibnamefont {Pagliara}}, \ and\ \bibinfo {author} {\bibfnamefont {D.}~\bibnamefont {Pigato}},\ }\href {\doibase 10.1103/PhysRevC.90.065809} {\bibfield  {journal} {\bibinfo  {journal} {Phys. Rev. C}\ }\textbf {\bibinfo {volume} {90}},\ \bibinfo {pages} {065809} (\bibinfo {year} {2014})},\ \Eprint {http://arxiv.org/abs/1407.2843} {arXiv:1407.2843 [astro-ph.SR]} \BibitemShut {NoStop}%
\bibitem [{\citenamefont {Malik}\ and\ \citenamefont {Provid\^encia}(2022)}]{Malik:2022jqc}%
  \BibitemOpen
  \bibfield  {author} {\bibinfo {author} {\bibfnamefont {T.}~\bibnamefont {Malik}}\ and\ \bibinfo {author} {\bibfnamefont {C.}~\bibnamefont {Provid\^encia}},\ }\href {\doibase 10.1103/PhysRevD.106.063024} {\bibfield  {journal} {\bibinfo  {journal} {Phys. Rev. D}\ }\textbf {\bibinfo {volume} {106}},\ \bibinfo {pages} {063024} (\bibinfo {year} {2022})},\ \Eprint {http://arxiv.org/abs/2205.15843} {arXiv:2205.15843 [nucl-th]} \BibitemShut {NoStop}%
\bibitem [{\citenamefont {Sun}\ \emph {et~al.}(2023)\citenamefont {Sun}, \citenamefont {Miao}, \citenamefont {Sun},\ and\ \citenamefont {Li}}]{Sun:2023}%
  \BibitemOpen
  \bibfield  {author} {\bibinfo {author} {\bibfnamefont {X.}~\bibnamefont {Sun}}, \bibinfo {author} {\bibfnamefont {Z.}~\bibnamefont {Miao}}, \bibinfo {author} {\bibfnamefont {B.}~\bibnamefont {Sun}}, \ and\ \bibinfo {author} {\bibfnamefont {A.}~\bibnamefont {Li}},\ }\href {\doibase 10.3847/1538-4357/ac9d9a} {\bibfield  {journal} {\bibinfo  {journal} {The Astrophysical Journal}\ }\textbf {\bibinfo {volume} {942}},\ \bibinfo {pages} {55} (\bibinfo {year} {2023})}\BibitemShut {NoStop}%
\bibitem [{\citenamefont {Ghosh}\ \emph {et~al.}(2022)\citenamefont {Ghosh}, \citenamefont {Pradhan}, \citenamefont {Chatterjee},\ and\ \citenamefont {Schaffner-Bielich}}]{Suprovo:2022}%
  \BibitemOpen
  \bibfield  {author} {\bibinfo {author} {\bibfnamefont {S.}~\bibnamefont {Ghosh}}, \bibinfo {author} {\bibfnamefont {B.~K.}\ \bibnamefont {Pradhan}}, \bibinfo {author} {\bibfnamefont {D.}~\bibnamefont {Chatterjee}}, \ and\ \bibinfo {author} {\bibfnamefont {J.}~\bibnamefont {Schaffner-Bielich}},\ }\href {\doibase 10.3389/fspas.2022.864294} {\bibfield  {journal} {\bibinfo  {journal} {Frontiers in Astronomy and Space Sciences}\ }\textbf {\bibinfo {volume} {9}} (\bibinfo {year} {2022}),\ 10.3389/fspas.2022.864294}\BibitemShut {NoStop}%
\bibitem [{\citenamefont {Huang}\ \emph {et~al.}(2024{\natexlab{a}})\citenamefont {Huang}, \citenamefont {Tolos}, \citenamefont {Provid\^encia},\ and\ \citenamefont {Watts}}]{Huang:2024rvj}%
  \BibitemOpen
  \bibfield  {author} {\bibinfo {author} {\bibfnamefont {C.}~\bibnamefont {Huang}}, \bibinfo {author} {\bibfnamefont {L.}~\bibnamefont {Tolos}}, \bibinfo {author} {\bibfnamefont {C.}~\bibnamefont {Provid\^encia}}, \ and\ \bibinfo {author} {\bibfnamefont {A.}~\bibnamefont {Watts}},\ }\href {\doibase 10.1093/mnras/stae2792} {\  (\bibinfo {year} {2024}{\natexlab{a}}),\ 10.1093/mnras/stae2792},\ \Eprint {http://arxiv.org/abs/2410.14572} {arXiv:2410.14572 [astro-ph.HE]} \BibitemShut {NoStop}%
\bibitem [{\citenamefont {Parmar}\ \emph {et~al.}(2025)\citenamefont {Parmar}, \citenamefont {Thapa}, \citenamefont {Sinha},\ and\ \citenamefont {Bombaci}}]{Parmar:2025csx}%
  \BibitemOpen
  \bibfield  {author} {\bibinfo {author} {\bibfnamefont {V.}~\bibnamefont {Parmar}}, \bibinfo {author} {\bibfnamefont {V.~B.}\ \bibnamefont {Thapa}}, \bibinfo {author} {\bibfnamefont {M.}~\bibnamefont {Sinha}}, \ and\ \bibinfo {author} {\bibfnamefont {I.}~\bibnamefont {Bombaci}},\ }\href@noop {} {\  (\bibinfo {year} {2025})},\ \Eprint {http://arxiv.org/abs/2503.07256} {arXiv:2503.07256 [astro-ph.HE]} \BibitemShut {NoStop}%
\bibitem [{\citenamefont {Glendenning}\ and\ \citenamefont {Moszkowski}(1991)}]{PhysRevLett.67.2414}%
  \BibitemOpen
  \bibfield  {author} {\bibinfo {author} {\bibfnamefont {N.~K.}\ \bibnamefont {Glendenning}}\ and\ \bibinfo {author} {\bibfnamefont {S.~A.}\ \bibnamefont {Moszkowski}},\ }\href {\doibase 10.1103/PhysRevLett.67.2414} {\bibfield  {journal} {\bibinfo  {journal} {Phys. Rev. Lett.}\ }\textbf {\bibinfo {volume} {67}},\ \bibinfo {pages} {2414} (\bibinfo {year} {1991})}\BibitemShut {NoStop}%
\bibitem [{\citenamefont {Glendenning}(2000)}]{Glendenning2000}%
  \BibitemOpen
  \bibfield  {author} {\bibinfo {author} {\bibfnamefont {N.~K.}\ \bibnamefont {Glendenning}},\ }\href@noop {} {\emph {\bibinfo {title} {Compact Star}}}\ (\bibinfo  {publisher} {Springer-Verlag},\ \bibinfo {address} {New York},\ \bibinfo {year} {2000})\BibitemShut {NoStop}%
\bibitem [{\citenamefont {Typel}\ and\ \citenamefont {Wolter}(1999)}]{Typel:1999yq}%
  \BibitemOpen
  \bibfield  {author} {\bibinfo {author} {\bibfnamefont {S.}~\bibnamefont {Typel}}\ and\ \bibinfo {author} {\bibfnamefont {H.~H.}\ \bibnamefont {Wolter}},\ }\href {\doibase 10.1016/S0375-9474(99)00310-3} {\bibfield  {journal} {\bibinfo  {journal} {Nucl. Phys. A}\ }\textbf {\bibinfo {volume} {656}},\ \bibinfo {pages} {331} (\bibinfo {year} {1999})}\BibitemShut {NoStop}%
\bibitem [{\citenamefont {Huang}\ \emph {et~al.}(2024{\natexlab{b}})\citenamefont {Huang}, \citenamefont {Raaijmakers}, \citenamefont {Watts}, \citenamefont {Tolos},\ and\ \citenamefont {Providência}}]{CHuang:2024}%
  \BibitemOpen
  \bibfield  {author} {\bibinfo {author} {\bibfnamefont {C.}~\bibnamefont {Huang}}, \bibinfo {author} {\bibfnamefont {G.}~\bibnamefont {Raaijmakers}}, \bibinfo {author} {\bibfnamefont {A.~L.}\ \bibnamefont {Watts}}, \bibinfo {author} {\bibfnamefont {L.}~\bibnamefont {Tolos}}, \ and\ \bibinfo {author} {\bibfnamefont {C.}~\bibnamefont {Providência}},\ }\href {https://arxiv.org/abs/2303.17518} {\enquote {\bibinfo {title} {Constraining a relativistic mean field model using neutron star mass-radius measurements i: Nucleonic models},}\ } (\bibinfo {year} {2024}{\natexlab{b}}),\ \Eprint {http://arxiv.org/abs/2303.17518} {arXiv:2303.17518 [astro-ph.HE]} \BibitemShut {NoStop}%
\bibitem [{\citenamefont {Wu}\ \emph {et~al.}(2021)\citenamefont {Wu}, \citenamefont {Bao}, \citenamefont {Shen},\ and\ \citenamefont {Xu}}]{PhysRevC.104.015802}%
  \BibitemOpen
  \bibfield  {author} {\bibinfo {author} {\bibfnamefont {X.}~\bibnamefont {Wu}}, \bibinfo {author} {\bibfnamefont {S.}~\bibnamefont {Bao}}, \bibinfo {author} {\bibfnamefont {H.}~\bibnamefont {Shen}}, \ and\ \bibinfo {author} {\bibfnamefont {R.}~\bibnamefont {Xu}},\ }\href {\doibase 10.1103/PhysRevC.104.015802} {\bibfield  {journal} {\bibinfo  {journal} {Phys. Rev. C}\ }\textbf {\bibinfo {volume} {104}},\ \bibinfo {pages} {015802} (\bibinfo {year} {2021})}\BibitemShut {NoStop}%
\bibitem [{\citenamefont {Baym}\ \emph {et~al.}(1971)\citenamefont {Baym}, \citenamefont {Pethick},\ and\ \citenamefont {Sutherland}}]{Baym:1971pw}%
  \BibitemOpen
  \bibfield  {author} {\bibinfo {author} {\bibfnamefont {G.}~\bibnamefont {Baym}}, \bibinfo {author} {\bibfnamefont {C.}~\bibnamefont {Pethick}}, \ and\ \bibinfo {author} {\bibfnamefont {P.}~\bibnamefont {Sutherland}},\ }\href {\doibase 10.1086/151216} {\bibfield  {journal} {\bibinfo  {journal} {Astrophys. J.}\ }\textbf {\bibinfo {volume} {170}},\ \bibinfo {pages} {299} (\bibinfo {year} {1971})}\BibitemShut {NoStop}%
\bibitem [{\citenamefont {Abbott}\ \emph {et~al.}(2017)\citenamefont {Abbott}, \citenamefont {Abbott}, \citenamefont {Abbott} \emph {et~al.}}]{Abbott:2017}%
  \BibitemOpen
  \bibfield  {author} {\bibinfo {author} {\bibfnamefont {B.~P.}\ \bibnamefont {Abbott}}, \bibinfo {author} {\bibfnamefont {R.}~\bibnamefont {Abbott}}, \bibinfo {author} {\bibfnamefont {T.~D.}\ \bibnamefont {Abbott}},  \emph {et~al.} (\bibinfo {collaboration} {LIGO Scientific Collaboration and Virgo Collaboration}),\ }\href {\doibase 10.1103/PhysRevLett.119.161101} {\bibfield  {journal} {\bibinfo  {journal} {\prl}\ }\textbf {\bibinfo {volume} {119}},\ \bibinfo {eid} {161101} (\bibinfo {year} {2017})},\ \Eprint {http://arxiv.org/abs/1710.05832} {arXiv:1710.05832 [gr-qc]} \BibitemShut {NoStop}%
\bibitem [{\citenamefont {Moustakidis}\ \emph {et~al.}(2017)\citenamefont {Moustakidis}, \citenamefont {Gaitanos}, \citenamefont {Margaritis},\ and\ \citenamefont {Lalazissis}}]{Moustakidis:2017}%
  \BibitemOpen
  \bibfield  {author} {\bibinfo {author} {\bibfnamefont {C.~C.}\ \bibnamefont {Moustakidis}}, \bibinfo {author} {\bibfnamefont {T.}~\bibnamefont {Gaitanos}}, \bibinfo {author} {\bibfnamefont {C.}~\bibnamefont {Margaritis}}, \ and\ \bibinfo {author} {\bibfnamefont {G.~A.}\ \bibnamefont {Lalazissis}},\ }\href {\doibase 10.1103/PhysRevC.95.045801} {\bibfield  {journal} {\bibinfo  {journal} {Phys. Rev. C}\ }\textbf {\bibinfo {volume} {95}},\ \bibinfo {pages} {045801} (\bibinfo {year} {2017})}\BibitemShut {NoStop}%
\bibitem [{\citenamefont {Tews}\ \emph {et~al.}(2016)\citenamefont {Tews}, \citenamefont {Gandolfi}, \citenamefont {Gezerlis},\ and\ \citenamefont {Schwenk}}]{Tews:2016}%
  \BibitemOpen
  \bibfield  {author} {\bibinfo {author} {\bibfnamefont {I.}~\bibnamefont {Tews}}, \bibinfo {author} {\bibfnamefont {S.}~\bibnamefont {Gandolfi}}, \bibinfo {author} {\bibfnamefont {A.}~\bibnamefont {Gezerlis}}, \ and\ \bibinfo {author} {\bibfnamefont {A.}~\bibnamefont {Schwenk}},\ }\href {\doibase 10.1103/PhysRevC.93.024305} {\bibfield  {journal} {\bibinfo  {journal} {Phys. Rev. C}\ }\textbf {\bibinfo {volume} {93}},\ \bibinfo {pages} {024305} (\bibinfo {year} {2016})}\BibitemShut {NoStop}%
\bibitem [{\citenamefont {Drischler}\ \emph {et~al.}(2016)\citenamefont {Drischler}, \citenamefont {Hebeler},\ and\ \citenamefont {Schwenk}}]{Drischler:2016}%
  \BibitemOpen
  \bibfield  {author} {\bibinfo {author} {\bibfnamefont {C.}~\bibnamefont {Drischler}}, \bibinfo {author} {\bibfnamefont {K.}~\bibnamefont {Hebeler}}, \ and\ \bibinfo {author} {\bibfnamefont {A.}~\bibnamefont {Schwenk}},\ }\href {\doibase 10.1103/PhysRevC.93.054314} {\bibfield  {journal} {\bibinfo  {journal} {Phys. Rev. C}\ }\textbf {\bibinfo {volume} {93}},\ \bibinfo {pages} {054314} (\bibinfo {year} {2016})}\BibitemShut {NoStop}%
\bibitem [{\citenamefont {Drischler}\ \emph {et~al.}(2020{\natexlab{b}})\citenamefont {Drischler}, \citenamefont {Melendez}, \citenamefont {Furnstahl},\ and\ \citenamefont {Phillips}}]{Drischler:2020xEFT}%
  \BibitemOpen
  \bibfield  {author} {\bibinfo {author} {\bibfnamefont {C.}~\bibnamefont {Drischler}}, \bibinfo {author} {\bibfnamefont {J.~A.}\ \bibnamefont {Melendez}}, \bibinfo {author} {\bibfnamefont {R.~J.}\ \bibnamefont {Furnstahl}}, \ and\ \bibinfo {author} {\bibfnamefont {D.~R.}\ \bibnamefont {Phillips}},\ }\href {\doibase 10.1103/PhysRevC.102.054315} {\bibfield  {journal} {\bibinfo  {journal} {Phys. Rev. C}\ }\textbf {\bibinfo {volume} {102}},\ \bibinfo {pages} {054315} (\bibinfo {year} {2020}{\natexlab{b}})}\BibitemShut {NoStop}%
\bibitem [{\citenamefont {Demorest}\ \emph {et~al.}(2010)\citenamefont {Demorest}, \citenamefont {Pennucci}, \citenamefont {Ransom}, \citenamefont {Roberts},\ and\ \citenamefont {Hessels}}]{Demorest:2010}%
  \BibitemOpen
  \bibfield  {author} {\bibinfo {author} {\bibfnamefont {P.}~\bibnamefont {Demorest}}, \bibinfo {author} {\bibfnamefont {T.}~\bibnamefont {Pennucci}}, \bibinfo {author} {\bibfnamefont {S.}~\bibnamefont {Ransom}}, \bibinfo {author} {\bibfnamefont {M.}~\bibnamefont {Roberts}}, \ and\ \bibinfo {author} {\bibfnamefont {J.}~\bibnamefont {Hessels}},\ }\href {\doibase 10.1038/nature09466} {\bibfield  {journal} {\bibinfo  {journal} {Nature}\ }\textbf {\bibinfo {volume} {467}},\ \bibinfo {pages} {1081} (\bibinfo {year} {2010})},\ \Eprint {http://arxiv.org/abs/1010.5788} {arXiv:1010.5788 [astro-ph.HE]} \BibitemShut {NoStop}%
\bibitem [{\citenamefont {Fonseca}\ \emph {et~al.}(2016)\citenamefont {Fonseca} \emph {et~al.}}]{Fonseca:2016}%
  \BibitemOpen
  \bibfield  {author} {\bibinfo {author} {\bibfnamefont {E.}~\bibnamefont {Fonseca}} \emph {et~al.},\ }\href {\doibase 10.3847/0004-637X/832/2/167} {\bibfield  {journal} {\bibinfo  {journal} {Astrophys. J.}\ }\textbf {\bibinfo {volume} {832}},\ \bibinfo {pages} {167} (\bibinfo {year} {2016})},\ \Eprint {http://arxiv.org/abs/1603.00545} {arXiv:1603.00545 [astro-ph.HE]} \BibitemShut {NoStop}%
\bibitem [{\citenamefont {Arzoumanian}\ \emph {et~al.}(2018)\citenamefont {Arzoumanian} \emph {et~al.}}]{Arzoumanian:2018}%
  \BibitemOpen
  \bibfield  {author} {\bibinfo {author} {\bibfnamefont {Z.}~\bibnamefont {Arzoumanian}} \emph {et~al.} (\bibinfo {collaboration} {NANOGrav}),\ }\href {\doibase 10.3847/1538-4365/aab5b0} {\bibfield  {journal} {\bibinfo  {journal} {Astrophys. J. Suppl.}\ }\textbf {\bibinfo {volume} {235}},\ \bibinfo {pages} {37} (\bibinfo {year} {2018})},\ \Eprint {http://arxiv.org/abs/1801.01837} {arXiv:1801.01837 [astro-ph.HE]} \BibitemShut {NoStop}%
\bibitem [{\citenamefont {Alam}\ \emph {et~al.}(2021)\citenamefont {Alam}, \citenamefont {Arzoumanian}, \citenamefont {Baker}, \citenamefont {Blumer} \emph {et~al.}}]{Alam:2021}%
  \BibitemOpen
  \bibfield  {author} {\bibinfo {author} {\bibfnamefont {M.}~\bibnamefont {Alam}}, \bibinfo {author} {\bibfnamefont {Z.}~\bibnamefont {Arzoumanian}}, \bibinfo {author} {\bibfnamefont {P.}~\bibnamefont {Baker}}, \bibinfo {author} {\bibfnamefont {H.}~\bibnamefont {Blumer}},  \emph {et~al.},\ }\href {\doibase 10.3847/1538-4365/abc6a0} {\bibfield  {journal} {\bibinfo  {journal} {Astrophysical Journal Supplement Series}\ }\textbf {\bibinfo {volume} {252}} (\bibinfo {year} {2021}),\ 10.3847/1538-4365/abc6a0}\BibitemShut {NoStop}%
\bibitem [{\citenamefont {Agazie}\ \emph {et~al.}(2023)\citenamefont {Agazie}, \citenamefont {Alam}, \citenamefont {Anumarlapudi}, \citenamefont {Archibald}, \citenamefont {Collaboration} \emph {et~al.}}]{Agazie:2023}%
  \BibitemOpen
  \bibfield  {author} {\bibinfo {author} {\bibfnamefont {G.}~\bibnamefont {Agazie}}, \bibinfo {author} {\bibfnamefont {M.~F.}\ \bibnamefont {Alam}}, \bibinfo {author} {\bibfnamefont {A.}~\bibnamefont {Anumarlapudi}}, \bibinfo {author} {\bibfnamefont {A.~M.}\ \bibnamefont {Archibald}}, \bibinfo {author} {\bibfnamefont {T.~N.}\ \bibnamefont {Collaboration}},  \emph {et~al.},\ }\href {\doibase 10.3847/2041-8213/acda9a} {\bibfield  {journal} {\bibinfo  {journal} {The Astrophysical Journal Letters}\ }\textbf {\bibinfo {volume} {951}},\ \bibinfo {pages} {L9} (\bibinfo {year} {2023})}\BibitemShut {NoStop}%
\bibitem [{\citenamefont {Antoniadis}\ \emph {et~al.}(2013)\citenamefont {Antoniadis}, \citenamefont {Freire}, \citenamefont {Wex}, \citenamefont {Tauris} \emph {et~al.}}]{Antoniadis:2013}%
  \BibitemOpen
  \bibfield  {author} {\bibinfo {author} {\bibfnamefont {J.}~\bibnamefont {Antoniadis}}, \bibinfo {author} {\bibfnamefont {P.~C.}\ \bibnamefont {Freire}}, \bibinfo {author} {\bibfnamefont {N.}~\bibnamefont {Wex}}, \bibinfo {author} {\bibfnamefont {T.~M.}\ \bibnamefont {Tauris}},  \emph {et~al.},\ }\href {\doibase 10.1126/science.1233232} {\bibfield  {journal} {\bibinfo  {journal} {Science}\ }\textbf {\bibinfo {volume} {340}},\ \bibinfo {pages} {6131} (\bibinfo {year} {2013})}\BibitemShut {NoStop}%
\bibitem [{\citenamefont {Cromartie}\ \emph {et~al.}(2019)\citenamefont {Cromartie}, \citenamefont {Fonseca}, \citenamefont {Ransom}, \citenamefont {Demorest} \emph {et~al.}}]{Cromartie:2019}%
  \BibitemOpen
  \bibfield  {author} {\bibinfo {author} {\bibfnamefont {H.~T.}\ \bibnamefont {Cromartie}}, \bibinfo {author} {\bibfnamefont {E.}~\bibnamefont {Fonseca}}, \bibinfo {author} {\bibfnamefont {S.~M.}\ \bibnamefont {Ransom}}, \bibinfo {author} {\bibfnamefont {P.~B.}\ \bibnamefont {Demorest}},  \emph {et~al.},\ }\href {\doibase 10.1038/s41550-019-0880-2} {\bibfield  {journal} {\bibinfo  {journal} {Nature Astron.}\ }\textbf {\bibinfo {volume} {4}},\ \bibinfo {pages} {72} (\bibinfo {year} {2019})},\ \Eprint {http://arxiv.org/abs/1904.06759} {arXiv:1904.06759 [astro-ph.HE]} \BibitemShut {NoStop}%
\bibitem [{\citenamefont {{Fonseca}}\ \emph {et~al.}()\citenamefont {{Fonseca}}, \citenamefont {{Cromartie}}, \citenamefont {{Pennucci}}, \citenamefont {{Ray}} \emph {et~al.}}]{Fonseca:2021}%
  \BibitemOpen
  \bibfield  {author} {\bibinfo {author} {\bibfnamefont {E.}~\bibnamefont {{Fonseca}}}, \bibinfo {author} {\bibfnamefont {H.~T.}\ \bibnamefont {{Cromartie}}}, \bibinfo {author} {\bibfnamefont {T.~T.}\ \bibnamefont {{Pennucci}}}, \bibinfo {author} {\bibfnamefont {P.~S.}\ \bibnamefont {{Ray}}},  \emph {et~al.},\ }\href@noop {} {\ }\BibitemShut {NoStop}%
\bibitem [{\citenamefont {Linares}\ \emph {et~al.}(2018)\citenamefont {Linares}, \citenamefont {Shahbaz},\ and\ \citenamefont {Casares}}]{Linares:2018}%
  \BibitemOpen
  \bibfield  {author} {\bibinfo {author} {\bibfnamefont {M.}~\bibnamefont {Linares}}, \bibinfo {author} {\bibfnamefont {T.}~\bibnamefont {Shahbaz}}, \ and\ \bibinfo {author} {\bibfnamefont {J.}~\bibnamefont {Casares}},\ }\href {\doibase 10.3847/1538-4357/aabde6} {\bibfield  {journal} {\bibinfo  {journal} {Astrophys. J.}\ }\textbf {\bibinfo {volume} {859}},\ \bibinfo {pages} {54} (\bibinfo {year} {2018})},\ \Eprint {http://arxiv.org/abs/1805.08799} {arXiv:1805.08799 [astro-ph.HE]} \BibitemShut {NoStop}%
\bibitem [{\citenamefont {Huang}\ \emph {et~al.}(2024{\natexlab{c}})\citenamefont {Huang}, \citenamefont {Raaijmakers}, \citenamefont {Watts}, \citenamefont {Tolos},\ and\ \citenamefont {Provid\^encia}}]{Huang:2023grj}%
  \BibitemOpen
  \bibfield  {author} {\bibinfo {author} {\bibfnamefont {C.}~\bibnamefont {Huang}}, \bibinfo {author} {\bibfnamefont {G.}~\bibnamefont {Raaijmakers}}, \bibinfo {author} {\bibfnamefont {A.~L.}\ \bibnamefont {Watts}}, \bibinfo {author} {\bibfnamefont {L.}~\bibnamefont {Tolos}}, \ and\ \bibinfo {author} {\bibfnamefont {C.}~\bibnamefont {Provid\^encia}},\ }\href {\doibase 10.1093/mnras/stae844} {\bibfield  {journal} {\bibinfo  {journal} {Mon. Not. Roy. Astron. Soc.}\ }\textbf {\bibinfo {volume} {529}},\ \bibinfo {pages} {4650} (\bibinfo {year} {2024}{\natexlab{c}})},\ \Eprint {http://arxiv.org/abs/2303.17518} {arXiv:2303.17518 [astro-ph.HE]} \BibitemShut {NoStop}%
\bibitem [{\citenamefont {Vinciguerra}\ \emph {et~al.}(2024)\citenamefont {Vinciguerra}, \citenamefont {Salmi}, \citenamefont {Watts}, \citenamefont {Choudhury}, \citenamefont {Riley}, \citenamefont {Ray}, \citenamefont {Bogdanov}, \citenamefont {Kini}, \citenamefont {Guillot}, \citenamefont {Chakrabarty}, \citenamefont {Ho}, \citenamefont {Huppenkothen}, \citenamefont {Morsink}, \citenamefont {Wadiasingh},\ and\ \citenamefont {Wolff}}]{Vinciguerra_2024}%
  \BibitemOpen
  \bibfield  {author} {\bibinfo {author} {\bibfnamefont {S.}~\bibnamefont {Vinciguerra}}, \bibinfo {author} {\bibfnamefont {T.}~\bibnamefont {Salmi}}, \bibinfo {author} {\bibfnamefont {A.~L.}\ \bibnamefont {Watts}}, \bibinfo {author} {\bibfnamefont {D.}~\bibnamefont {Choudhury}}, \bibinfo {author} {\bibfnamefont {T.~E.}\ \bibnamefont {Riley}}, \bibinfo {author} {\bibfnamefont {P.~S.}\ \bibnamefont {Ray}}, \bibinfo {author} {\bibfnamefont {S.}~\bibnamefont {Bogdanov}}, \bibinfo {author} {\bibfnamefont {Y.}~\bibnamefont {Kini}}, \bibinfo {author} {\bibfnamefont {S.}~\bibnamefont {Guillot}}, \bibinfo {author} {\bibfnamefont {D.}~\bibnamefont {Chakrabarty}}, \bibinfo {author} {\bibfnamefont {W.~C.~G.}\ \bibnamefont {Ho}}, \bibinfo {author} {\bibfnamefont {D.}~\bibnamefont {Huppenkothen}}, \bibinfo {author} {\bibfnamefont {S.~M.}\ \bibnamefont {Morsink}}, \bibinfo {author} {\bibfnamefont {Z.}~\bibnamefont {Wadiasingh}}, \ and\ \bibinfo {author} {\bibfnamefont {M.~T.}\ \bibnamefont {Wolff}},\ }\href {\doibase
  10.3847/1538-4357/acfb83} {\bibfield  {journal} {\bibinfo  {journal} {The Astrophysical Journal}\ }\textbf {\bibinfo {volume} {961}},\ \bibinfo {pages} {62} (\bibinfo {year} {2024})}\BibitemShut {NoStop}%
\bibitem [{\citenamefont {Miller}\ \emph {et~al.}(2021)\citenamefont {Miller}, \citenamefont {Lamb}, \citenamefont {Dittmann}, \citenamefont {Bogdanov}, \citenamefont {Arzoumanian}, \citenamefont {Gendreau}, \citenamefont {Guillot}, \citenamefont {Ho}, \citenamefont {Lattimer}, \citenamefont {Loewenstein}, \citenamefont {Morsink}, \citenamefont {Ray}, \citenamefont {Wolff}, \citenamefont {Baker}, \citenamefont {Cazeau}, \citenamefont {Manthripragada}, \citenamefont {Markwardt}, \citenamefont {Okajima}, \citenamefont {Pollard}, \citenamefont {Cognard}, \citenamefont {Cromartie}, \citenamefont {Fonseca}, \citenamefont {Guillemot}, \citenamefont {Kerr}, \citenamefont {Parthasarathy}, \citenamefont {Pennucci}, \citenamefont {Ransom},\ and\ \citenamefont {Stairs}}]{Miller_2021}%
  \BibitemOpen
  \bibfield  {author} {\bibinfo {author} {\bibfnamefont {M.~C.}\ \bibnamefont {Miller}}, \bibinfo {author} {\bibfnamefont {F.~K.}\ \bibnamefont {Lamb}}, \bibinfo {author} {\bibfnamefont {A.~J.}\ \bibnamefont {Dittmann}}, \bibinfo {author} {\bibfnamefont {S.}~\bibnamefont {Bogdanov}}, \bibinfo {author} {\bibfnamefont {Z.}~\bibnamefont {Arzoumanian}}, \bibinfo {author} {\bibfnamefont {K.~C.}\ \bibnamefont {Gendreau}}, \bibinfo {author} {\bibfnamefont {S.}~\bibnamefont {Guillot}}, \bibinfo {author} {\bibfnamefont {W.~C.~G.}\ \bibnamefont {Ho}}, \bibinfo {author} {\bibfnamefont {J.~M.}\ \bibnamefont {Lattimer}}, \bibinfo {author} {\bibfnamefont {M.}~\bibnamefont {Loewenstein}}, \bibinfo {author} {\bibfnamefont {S.~M.}\ \bibnamefont {Morsink}}, \bibinfo {author} {\bibfnamefont {P.~S.}\ \bibnamefont {Ray}}, \bibinfo {author} {\bibfnamefont {M.~T.}\ \bibnamefont {Wolff}}, \bibinfo {author} {\bibfnamefont {C.~L.}\ \bibnamefont {Baker}}, \bibinfo {author} {\bibfnamefont {T.}~\bibnamefont {Cazeau}}, \bibinfo {author}
  {\bibfnamefont {S.}~\bibnamefont {Manthripragada}}, \bibinfo {author} {\bibfnamefont {C.~B.}\ \bibnamefont {Markwardt}}, \bibinfo {author} {\bibfnamefont {T.}~\bibnamefont {Okajima}}, \bibinfo {author} {\bibfnamefont {S.}~\bibnamefont {Pollard}}, \bibinfo {author} {\bibfnamefont {I.}~\bibnamefont {Cognard}}, \bibinfo {author} {\bibfnamefont {H.~T.}\ \bibnamefont {Cromartie}}, \bibinfo {author} {\bibfnamefont {E.}~\bibnamefont {Fonseca}}, \bibinfo {author} {\bibfnamefont {L.}~\bibnamefont {Guillemot}}, \bibinfo {author} {\bibfnamefont {M.}~\bibnamefont {Kerr}}, \bibinfo {author} {\bibfnamefont {A.}~\bibnamefont {Parthasarathy}}, \bibinfo {author} {\bibfnamefont {T.~T.}\ \bibnamefont {Pennucci}}, \bibinfo {author} {\bibfnamefont {S.}~\bibnamefont {Ransom}}, \ and\ \bibinfo {author} {\bibfnamefont {I.}~\bibnamefont {Stairs}},\ }\href {\doibase 10.3847/2041-8213/ac089b} {\bibfield  {journal} {\bibinfo  {journal} {The Astrophysical Journal Letters}\ }\textbf {\bibinfo {volume} {918}},\ \bibinfo {pages} {L28}
  (\bibinfo {year} {2021})}\BibitemShut {NoStop}%
\bibitem [{\citenamefont {Fortin}\ \emph {et~al.}(2015)\citenamefont {Fortin}, \citenamefont {Zdunik}, \citenamefont {Haensel},\ and\ \citenamefont {Bejger}}]{Fortin:2014mya}%
  \BibitemOpen
  \bibfield  {author} {\bibinfo {author} {\bibfnamefont {M.}~\bibnamefont {Fortin}}, \bibinfo {author} {\bibfnamefont {J.~L.}\ \bibnamefont {Zdunik}}, \bibinfo {author} {\bibfnamefont {P.}~\bibnamefont {Haensel}}, \ and\ \bibinfo {author} {\bibfnamefont {M.}~\bibnamefont {Bejger}},\ }\href {\doibase 10.1051/0004-6361/201424800} {\bibfield  {journal} {\bibinfo  {journal} {Astron. Astrophys.}\ }\textbf {\bibinfo {volume} {576}},\ \bibinfo {pages} {A68} (\bibinfo {year} {2015})},\ \Eprint {http://arxiv.org/abs/1408.3052} {arXiv:1408.3052 [astro-ph.SR]} \BibitemShut {NoStop}%
\bibitem [{\citenamefont {Lonardoni}\ \emph {et~al.}(2015)\citenamefont {Lonardoni}, \citenamefont {Lovato}, \citenamefont {Gandolfi},\ and\ \citenamefont {Pederiva}}]{Lonardoni:2014bwa}%
  \BibitemOpen
  \bibfield  {author} {\bibinfo {author} {\bibfnamefont {D.}~\bibnamefont {Lonardoni}}, \bibinfo {author} {\bibfnamefont {A.}~\bibnamefont {Lovato}}, \bibinfo {author} {\bibfnamefont {S.}~\bibnamefont {Gandolfi}}, \ and\ \bibinfo {author} {\bibfnamefont {F.}~\bibnamefont {Pederiva}},\ }\href {\doibase 10.1103/PhysRevLett.114.092301} {\bibfield  {journal} {\bibinfo  {journal} {Phys. Rev. Lett.}\ }\textbf {\bibinfo {volume} {114}},\ \bibinfo {pages} {092301} (\bibinfo {year} {2015})},\ \Eprint {http://arxiv.org/abs/1407.4448} {arXiv:1407.4448 [nucl-th]} \BibitemShut {NoStop}%
\bibitem [{\citenamefont {Espinoza}\ \emph {et~al.}(2011)\citenamefont {Espinoza}, \citenamefont {Lyne}, \citenamefont {Stappers},\ and\ \citenamefont {Kramer}}]{Espinoza:2011}%
  \BibitemOpen
  \bibfield  {author} {\bibinfo {author} {\bibfnamefont {C.~M.}\ \bibnamefont {Espinoza}}, \bibinfo {author} {\bibfnamefont {A.~G.}\ \bibnamefont {Lyne}}, \bibinfo {author} {\bibfnamefont {B.~W.}\ \bibnamefont {Stappers}}, \ and\ \bibinfo {author} {\bibfnamefont {M.}~\bibnamefont {Kramer}},\ }\href {\doibase 10.1111/j.1365-2966.2011.18503.x} {\bibfield  {journal} {\bibinfo  {journal} {Monthly Notices of the Royal Astronomical Society}\ }\textbf {\bibinfo {volume} {414}},\ \bibinfo {pages} {1679} (\bibinfo {year} {2011})},\ \Eprint {http://arxiv.org/abs/https://academic.oup.com/mnras/article-pdf/414/2/1679/3015039/mnras0414-1679.pdf} {https://academic.oup.com/mnras/article-pdf/414/2/1679/3015039/mnras0414-1679.pdf} \BibitemShut {NoStop}%
\bibitem [{\citenamefont {Link}\ \emph {et~al.}(1999)\citenamefont {Link}, \citenamefont {Epstein},\ and\ \citenamefont {Lattimer}}]{Link:1999}%
  \BibitemOpen
  \bibfield  {author} {\bibinfo {author} {\bibfnamefont {B.}~\bibnamefont {Link}}, \bibinfo {author} {\bibfnamefont {R.~I.}\ \bibnamefont {Epstein}}, \ and\ \bibinfo {author} {\bibfnamefont {J.~M.}\ \bibnamefont {Lattimer}},\ }\href {\doibase 10.1103/PhysRevLett.83.3362} {\bibfield  {journal} {\bibinfo  {journal} {Phys. Rev. Lett.}\ }\textbf {\bibinfo {volume} {83}},\ \bibinfo {pages} {3362} (\bibinfo {year} {1999})}\BibitemShut {NoStop}%
\bibitem [{\citenamefont {Haskell}\ and\ \citenamefont {Melatos}(2015)}]{Haskell:2015}%
  \BibitemOpen
  \bibfield  {author} {\bibinfo {author} {\bibfnamefont {B.}~\bibnamefont {Haskell}}\ and\ \bibinfo {author} {\bibfnamefont {A.}~\bibnamefont {Melatos}},\ }\href {\doibase 10.1142/S0218271815300086} {\bibfield  {journal} {\bibinfo  {journal} {International Journal of Modern Physics D}\ }\textbf {\bibinfo {volume} {24}},\ \bibinfo {pages} {1530008} (\bibinfo {year} {2015})},\ \Eprint {http://arxiv.org/abs/https://doi.org/10.1142/S0218271815300086} {https://doi.org/10.1142/S0218271815300086} \BibitemShut {NoStop}%
\bibitem [{\citenamefont {Fortin}\ \emph {et~al.}(2016)\citenamefont {Fortin}, \citenamefont {Provid\^encia}, \citenamefont {Raduta}, \citenamefont {Gulminelli}, \citenamefont {Zdunik}, \citenamefont {Haensel},\ and\ \citenamefont {Bejger}}]{Fortin:2016}%
  \BibitemOpen
  \bibfield  {author} {\bibinfo {author} {\bibfnamefont {M.}~\bibnamefont {Fortin}}, \bibinfo {author} {\bibfnamefont {C.}~\bibnamefont {Provid\^encia}}, \bibinfo {author} {\bibfnamefont {A.~R.}\ \bibnamefont {Raduta}}, \bibinfo {author} {\bibfnamefont {F.}~\bibnamefont {Gulminelli}}, \bibinfo {author} {\bibfnamefont {J.~L.}\ \bibnamefont {Zdunik}}, \bibinfo {author} {\bibfnamefont {P.}~\bibnamefont {Haensel}}, \ and\ \bibinfo {author} {\bibfnamefont {M.}~\bibnamefont {Bejger}},\ }\href@noop {} {\bibfield  {journal} {\bibinfo  {journal} {Phys. Rev. C}\ }\textbf {\bibinfo {volume} {94}},\ \bibinfo {pages} {035804} (\bibinfo {year} {2016})}\BibitemShut {NoStop}%
\bibitem [{\citenamefont {Carreau}\ \emph {et~al.}(2019)\citenamefont {Carreau}, \citenamefont {Gulminelli},\ and\ \citenamefont {Margueron}}]{Carreau:2019}%
  \BibitemOpen
  \bibfield  {author} {\bibinfo {author} {\bibfnamefont {T.}~\bibnamefont {Carreau}}, \bibinfo {author} {\bibfnamefont {F.}~\bibnamefont {Gulminelli}}, \ and\ \bibinfo {author} {\bibfnamefont {J.}~\bibnamefont {Margueron}},\ }\href {\doibase 10.1140/epja/i2019-12884-1} {\bibfield  {journal} {\bibinfo  {journal} {Eur. Phys. J. A}\ }\textbf {\bibinfo {volume} {55}},\ \bibinfo {pages} {188} (\bibinfo {year} {2019})}\BibitemShut {NoStop}%
\bibitem [{\citenamefont {Lattimer}\ \emph {et~al.}(1991)\citenamefont {Lattimer}, \citenamefont {Pethick}, \citenamefont {Prakash},\ and\ \citenamefont {Haensel}}]{Lattimer:1991}%
  \BibitemOpen
  \bibfield  {author} {\bibinfo {author} {\bibfnamefont {J.~M.}\ \bibnamefont {Lattimer}}, \bibinfo {author} {\bibfnamefont {C.~J.}\ \bibnamefont {Pethick}}, \bibinfo {author} {\bibfnamefont {M.}~\bibnamefont {Prakash}}, \ and\ \bibinfo {author} {\bibfnamefont {P.}~\bibnamefont {Haensel}},\ }\href {\doibase 10.1103/PhysRevLett.66.2701} {\bibfield  {journal} {\bibinfo  {journal} {Phys. Rev. Lett.}\ }\textbf {\bibinfo {volume} {66}},\ \bibinfo {pages} {2701} (\bibinfo {year} {1991})}\BibitemShut {NoStop}%
\bibitem [{\citenamefont {Somasundaram}\ \emph {et~al.}(2021)\citenamefont {Somasundaram}, \citenamefont {Drischler}, \citenamefont {Tews},\ and\ \citenamefont {Margueron}}]{Somasundaram:2021}%
  \BibitemOpen
  \bibfield  {author} {\bibinfo {author} {\bibfnamefont {R.}~\bibnamefont {Somasundaram}}, \bibinfo {author} {\bibfnamefont {C.}~\bibnamefont {Drischler}}, \bibinfo {author} {\bibfnamefont {I.}~\bibnamefont {Tews}}, \ and\ \bibinfo {author} {\bibfnamefont {J.}~\bibnamefont {Margueron}},\ }\href {\doibase 10.1103/PhysRevC.103.045803} {\bibfield  {journal} {\bibinfo  {journal} {Phys. Rev. C}\ }\textbf {\bibinfo {volume} {103}},\ \bibinfo {pages} {045803} (\bibinfo {year} {2021})}\BibitemShut {NoStop}%
\bibitem [{\citenamefont {Akmal}\ \emph {et~al.}(1998)\citenamefont {Akmal}, \citenamefont {Pandharipande},\ and\ \citenamefont {Ravenhall}}]{APR:1998}%
  \BibitemOpen
  \bibfield  {author} {\bibinfo {author} {\bibfnamefont {A.}~\bibnamefont {Akmal}}, \bibinfo {author} {\bibfnamefont {V.~R.}\ \bibnamefont {Pandharipande}}, \ and\ \bibinfo {author} {\bibfnamefont {D.~G.}\ \bibnamefont {Ravenhall}},\ }\href {\doibase 10.1103/PhysRevC.58.1804} {\bibfield  {journal} {\bibinfo  {journal} {Phys. Rev. C}\ }\textbf {\bibinfo {volume} {58}},\ \bibinfo {pages} {1804} (\bibinfo {year} {1998})}\BibitemShut {NoStop}%
\bibitem [{\citenamefont {Vidaña}\ \emph {et~al.}(2024)\citenamefont {Vidaña}, \citenamefont {Margueron},\ and\ \citenamefont {Schulze}}]{Vidana:2024}%
  \BibitemOpen
  \bibfield  {author} {\bibinfo {author} {\bibfnamefont {I.}~\bibnamefont {Vidaña}}, \bibinfo {author} {\bibfnamefont {J.}~\bibnamefont {Margueron}}, \ and\ \bibinfo {author} {\bibfnamefont {H.-J.}\ \bibnamefont {Schulze}},\ }\href {\doibase 10.3390/universe10050226} {\bibfield  {journal} {\bibinfo  {journal} {Universe}\ }\textbf {\bibinfo {volume} {10}} (\bibinfo {year} {2024}),\ 10.3390/universe10050226}\BibitemShut {NoStop}%
\bibitem [{\citenamefont {Goriely}\ \emph {et~al.}(2013)\citenamefont {Goriely}, \citenamefont {Chamel},\ and\ \citenamefont {Pearson}}]{Goriely:2013}%
  \BibitemOpen
  \bibfield  {author} {\bibinfo {author} {\bibfnamefont {S.}~\bibnamefont {Goriely}}, \bibinfo {author} {\bibfnamefont {N.}~\bibnamefont {Chamel}}, \ and\ \bibinfo {author} {\bibfnamefont {J.~M.}\ \bibnamefont {Pearson}},\ }\href {\doibase 10.1103/PhysRevC.88.061302} {\bibfield  {journal} {\bibinfo  {journal} {Phys. Rev. C}\ }\textbf {\bibinfo {volume} {88}},\ \bibinfo {pages} {061302} (\bibinfo {year} {2013})}\BibitemShut {NoStop}%
\end{thebibliography}
%
\end{document}